\documentstyle[epsf,11pt]{article}
\newcommand{\be}{\begin{equation}}
\newcommand{\ee}{\end{equation}}

\newcommand{\cmod}[1] {|{#1}|^2}

\newcommand\ph{\varphi_{i,j,\rho}}
\newcommand\phl{\varphi_{l(i,j,\rho)}}
\newcommand\phr{\varphi_{r(i,j,\rho)}}
\newcommand\phd{\varphi_{d(i,j,\rho)}}

\newcommand\Wl{Wl_{i,j,\rho}}
\newcommand\Wr{Wr_{i,j,\rho}}
\newcommand\Wd{Wd_{i,j,\rho}}

\newcommand\pc{\varphi}
\newcommand\pr{\varphi_r}
\newcommand\pl{\varphi_l}
\newcommand\pd{\varphi_d}

\newcommand\si{\rho}

\begin{document}
\title{Self-trapped electron states in nanotubes\\
}

\author{
L. Bratek\thanks{e-mail address: Lukasz.Bratek@ifj.edu.pl},\,
\\
Department of Mathematical Sciences,University of Durham, \\
Durham DH1 3LE, UK\\
L. Brizhik\thanks{e-mail address: brizhik@bitp.kiev.ua},\,
A. Eremko\thanks{e-mail address: eremko@bitp.kiev.ua},\,
\\
Bogolyubov Institute for Theoretical Physics, 03143 Kyiv, Ukraine\\
B. Piette\thanks{e-mail address: B.M.A.G.Piette@durham.ac.uk},\,
M. Watson\thanks{e-mail address: m.j.watson@durham.ac.uk},\,
and
W. Zakrzewski\thanks{e-mail address: W.J.Zakrzewski@durham.ac.uk}
\\
Department of Mathematical Sciences,University of Durham, \\
Durham DH1 3LE, UK\\
}
\date{}
\maketitle

\begin{abstract}

We study numerically self-trapped (polaron) states of quasiparticles 
(electrons, holes or excitons) in a deformable nanotube formed by a hexagonal 
lattice, wrapped into a cylinder (carbon- and boron nitride-type nanotube 
structures). We present a Hamiltonian for such a system taking into account  
an electron-phonon interaction, 
and determine conditions under which the lowest energy states are polarons.
We compute a large class of numerical solutions of this model for a wide 
range of the parameters. We show
that at not too strong electron-phonon coupling, the system admits  
ring-like localized solutions wrapped around 
the nanotube (the charge carrier is localized along the nanotube axis and 
uniformly distributed with respect to 
the azimuthal coordinate). At stronger coupling,  
solutions are localized on very few lattice sites in both directions of the 
nanotube. The transition from one type solution
to the other one depends on the diameter of the nanotube. We show that for
the values of the
carbon nanotube parameters, the polarons  have a ring-like structure wrapped 
around the nanotube with a profile resembling that of the nonlinear 
Schr\"odinger soliton.

\end{abstract}

\section{Introduction}

Low-dimensional systems have been studied intensively during the last decades.
Some systems have been found to possess unusual properties which turn 
out to be important for their practical applications  in microelectronics and 
nanotechnologies. Conducting polymers, quasi-one-dimensional organic and 
inorganic compounds are good examples of such systems \cite{Heeger}. 
Recently, nanotubes have become the subject of many intensive experimental 
and theoretical studies. In particular,  
carbon \cite{SaiDrDr,DrDrEkl,Dai} and boron nitride \cite{Benny} nanotubes 
have been shown to be useful for miniaturised electronic, mechanical, 
electromechanical and optoelectronic devices.
Some of the intriguing physical properties of such systems are due to 
the electron-phonon interactions which are particularly important 
in low-dimensional systems. 

Although low-dimensional systems have been studied mainly within simplified 
single-chain (single-band) models 
resulting in the theoretical predictions of many striking
phenomena \cite{Froehlich,Peierls,Dav}, many of them, such as Peierls 
transitions, formation of charge density waves, soliton states, etc
have been verified experimentally in real substances. 
However, taking into account the more realistic complex structure of these 
substances is quite important for the proper description of their 
properties and can result in qualitatively new features. This is particularly 
true in the case of nanotubes 
who possess series of energy bands which  are determined by one-dimensional 
(1D) energy dispersion relations on 
the wave vector $k$ along the axis of the nanotube. It is well known that 
sufficiently long single wall carbon nanotubes (SWNT), 
in which half of the energy band states are occupied, possess properties of 
1D-metals or semiconductors depending on 
their diameter and chirality \cite{SaiDrDr,DrDrEkl,SFDrDr,MDWh}. 
 In fact, their properties can be modified in a controllable way by varying 
either the nanotube diameter and chirality
or  by doping nanotubes with some impurity atoms, molecules or compounds 
\cite{Duc}.

 In 1D systems the electron-phonon interactions can lead to the formation of 
stable self-trapped states with spontaneous 
symmetry-breaking distortions \cite{Dav,SSH}. Such soliton states are realized 
in quasi-1D organic 
and inorganic materials and can successfully describe various properties of 
conjugated polymers in the framework of the Su-Schrieffer-Heeger model 
\cite{SSH}.
The possibility for the formation of states with a spontaneously broken 
symmetry in carbon nanotubes has been discussed in \cite{MDWh,Kane,Chamon}. 
In particular, large polarons (or solitons) in nanotubes 
have recently been considered in \cite{Alves,Pryl}. In these models  the 
long-wave approximation has been used for the states close to the Fermi-level, 
which is equivalent to the continuum approximation. In this way, however,  
the essential details of the crystallographic
structure of the system are not taken into account.   

Here we study numerically the formation and properties of polaron
 states in  zigzag nanotubes using the semi-empirical 
tight-binding model in which we consider only the nearest-neighbour hopping 
interactions \cite{SaiDrDr}. The merit of this method 
is its simplicity, in spite of which it often gives a quantitative agreement 
with the 
experimental results. This has been demonstrated in \cite{SSH} 
for conjugated polymers  and for 
carbon nanotubes in \cite{SaiDrDr,DrDrEkl,SFDrDr,MDWh}.  
Moreover, we consider only the ground state of a single electron in a 
hexagonal nanotube.  
Such systems are already quite complicated and though such states in 1D 
chains have been studied intensively both analytically and numerically, 
little is known about them in the case of nanotubes.

\section{The Model}
We consider a model in which a two-dimensional  sheet of hexagonally bound 
equivalent atoms (graphene-like sheet) is rolled-up into a zigzag 
nanotube.
Calling $d$ the length of each side of the hexagons, $R$ the radius 
of the nanotube, $n$ the number of hexagonal cells wrapped around 
the nanotube, we define (cfr. Fig 1)
\begin{equation}
\alpha = 2\pi / n, \qquad a = d \sqrt{3}, \qquad b = d/2.
\end{equation}
\begin{figure}[htbp]
\unitlength1cm \hfil
\begin{picture}(8,8)
 \epsfxsize=8cm \epsffile{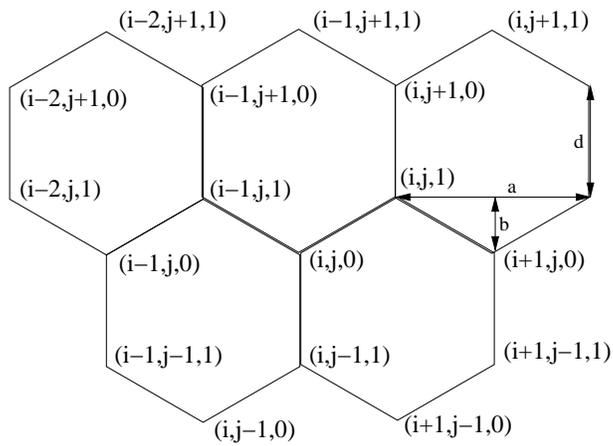}
\end{picture}
\caption{Hexagonal lattice, $j$ is assumed even.}
\end{figure}

The site labelling that we use is shown in Fig.1 and coincides with 
the conventional one \cite{SaiDrDr}. Here $j$ is an axial index, 
$i=1, ... , n$ is the index labelling the sites around the 
nanotube, and $\rho=1,2$ labels the two types of atoms 
(usually called  A and B): those who have a
nearest neighbour one site down ($\rho=0$) and those who have a nearest 
neighbour one site `up' ($\rho=1$) - see Fig.1.  Then the position of any 
nanotube lattice site, at its equilibrium, can be described by:  
\begin{eqnarray} 
\vec{R}^0_{i,j,\rho} = R \sin((i+{j+\rho\over 
2})\alpha)\vec{e}_x + R \cos((i+{j+\rho\over 2})\alpha) \vec{e}_y + 
{3j+\rho\over 2}d \vec{e}_z .  
\end{eqnarray} 

In Fig. 2 we present a drawing of the horizontal cross-section of the nanotube.
 Note that the parameters $R$, $a$ and $\alpha$ are connected by 
\begin{equation} 
a = 4 R \sin(\frac{\alpha}{4}).  
\end{equation}

\begin{figure}[htbp]
\unitlength1cm \hfil
\begin{picture}(8,8)
 \epsfxsize=8cm \epsffile{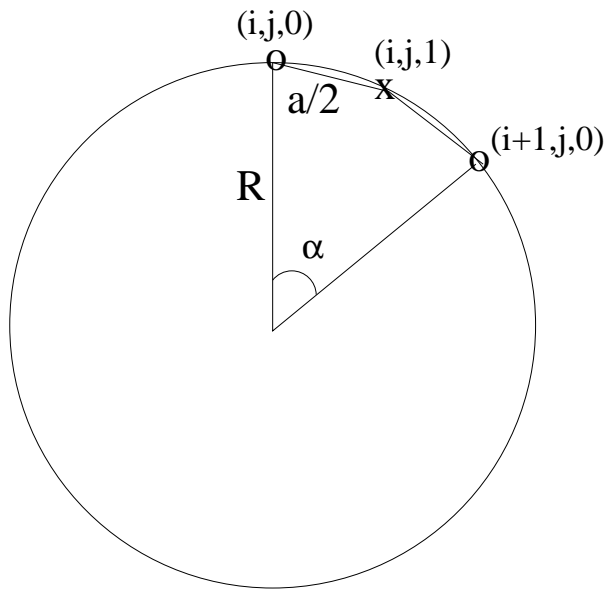}
\end{picture}
\caption{Cross section of the nanotube.}
\end{figure}

Next we  define
\begin{equation}
\vec{U}_{i,j,\rho} = \vec{u}_{i,j,\rho} + \vec{s}_{i,j,\rho}
                     + \vec{v}_{i,j,\rho}
\label{U_loc}		     
\end{equation}
as the displacement vector of the atoms from their equilibrium 
positions where $\vec{u}_{i,j,\rho}$ is tangent to the surface of 
the undeformed nanotube and perpendicular to the nanotube axis, 
$\vec{v}_{i,j,\rho}$ is tangent to this surface and parallel to 
the nanotube axis, and $\vec{s}_{i,j,\rho}$ is normal to the 
surface of the nanotube. Then, using Cartesian coordinates, we have
\begin{eqnarray}
&& \vec{u}_{i,j,\rho} =  u_{i,j,\rho}
      (\cos((i+(j+\rho)/2)\alpha) \vec{e_x}
       -\sin((i+(j+\rho)/2)\alpha) \vec{e_y}),\nonumber\\
&& \vec{s}_{i,j,\rho} =  s_{i,j,\rho}
      (\sin((i+(j+\rho)/2)\alpha) \vec{e_x} +
       \cos((i+(j+\rho)/2)\alpha) \vec{e_y}),\nonumber\\
&& \vec{v}_{i,j,\rho} =  v_{i,j,\rho} \vec{e_z},
\end{eqnarray}
where $u_{i,j,\rho}$, $s_{i,j,\rho}$ and $v_{i,j,\rho}$ are the corresponding 
amplitudes of displacements.
So the positions of the lattice sites are given by
\begin{equation}
\vec{R}_{i,j,\rho} = \vec{R}^0_{i,j,\rho} + \vec{U}_{i,j,\rho}.
\end{equation}

We also define the following lattice vectors connecting the atom $(i,j,\rho)$ 
with its three nearest neighbours $\delta(i,j,\rho)$ with $\delta = r,l,d$ 
for right ($r$), left ($l$) and down or up ($d$) neighbours: 
\begin{eqnarray}
\vec{D\delta}_{i,j,\rho} = \vec{R}_{\delta(i,j,\rho)}   - \vec{R}_{i,j,\rho} .
\end{eqnarray}
When $\vec{U}_{i,j,\rho}=0$  we add the upper index $0$ to 
 all quantities to indicate their values at the equilibrium position. Hence 
$\vec{D\delta}_{i,j,\rho}=\vec{D\delta}^0_{i,j,\rho}+\vec{d\delta}_{i,j,\rho}$ 
with $\vec{d\delta}_{i,j,\rho} = \vec{U}_{\delta(i,j,\rho)} - \vec{U}_{i,j,\rho}$. 
Note that
\begin{equation}
|\vec{Dr}^0_{i,j,\rho}| = |\vec{Dl}^0_{i,j,\rho}| =
|\vec{Dd}^0_{i,j,\rho}|= d.
\end{equation}

The potential energy of the lattice distortion includes central forces, which 
depend only on the distance between two sites. In the case of small 
displacement, {\it i.e} when $|\vec{d\delta}_{i,j,\rho}| \ll d$, the distance 
between lattice sites is approximately given by:
\begin{equation}
|\vec{D\delta}_{i,j,\rho}| \approx d + W\delta_{i,j,\rho},
\end{equation}
where 
\begin{equation}
 W\delta_{i,j,\rho} = \frac{\vec{d\delta}_{i,j,\rho} \cdot \vec{D\delta}^0_{i,j,\rho}}{d} 
\end{equation}
are the changes of distances between nearest neighbours due to the 
displacements of the sites.

As is well known, the central forces  between neighbouring sites do not provide 
lattice stability. Therefore, we introduce additional terms.
First, assuming that the force between two atoms is not purely central, 
in addition to $W\delta_{i,j,\rho}$ which
are invariant under translations, we introduce also the quantities 
$\Omega \delta_{i,j,\rho}$
which describe relative shifts of neighbouring atoms in a graphene sheet
 and which will be needed to remove the non-physical zero modes of the phonon 
Hamiltonian. The explicit expressions for these quantities are:
\begin{eqnarray}
&&\Omega r_{i,j,0} = {1\over2}\Big(
        \cos(\frac{\alpha}{4})(u_{i,j,1}-u_{i,j,0})+
         \sin(\frac{\alpha}{4})(s_{i,j,1}+s_{i,j,0})\Big)
         - \frac{\sqrt{3}}{2} (v_{i,j,1} - v_{i,j,0}),\nonumber\\
&&\Omega l_{i,j,0} = {1\over2}\Big(
        \cos(\frac{\alpha}{4})(u_{i,j,0}-u_{i-1,j,1})+
         \sin(\frac{\alpha}{4})(s_{i-1,j,1}+s_{i,j,0})\Big)
         - \frac{\sqrt{3}}{2} (v_{i-1,j,1} - v_{i,j,0}), \nonumber\\
&&\Omega d_{i,j,0} = -u_{i,j-1,1} + u_{i,j,0}\, \nonumber\\
&&\Omega r_{i,j,1} = \Omega r_{i,j,0}, \qquad \Omega l_{i,j,1} = \Omega l_{i+1,j,0}, 
   \qquad  \Omega d_{i,j,1} = \Omega d_{i,j+1,0}.
\end{eqnarray}

Next, as in \cite{WoMah,BEPZ_nc}, we introduce also the force of 
bond-bending between the atoms induced by the deviation of the valence 
angles from their  equilibrium values. For this we compute the solid angle 
spanned by the 3 lattice vectors located at a given site. 
Clearly, this angle is given by 
\begin{eqnarray} 
S_{i,j,\rho} &=& {(\vec{Dl}_{i,j,\rho} 
  \times \vec{Dr}_{i,j,\rho}) .  \vec{Dd}_{i,j,\rho}\over 
     |\vec{Dr}_{i,j,\rho}| |\vec{Dl}_{i,j,\rho}| |\vec{Dd}_{i,j,\rho}|}.
\end{eqnarray}
For small displacements we have
\begin{eqnarray}
S_{i,j,\rho} \approx&  S^0_{i,j,\rho} + \frac{\sqrt{3}}{2d} C_{i,j,\si}
\end{eqnarray}
where $S^0_{i,j,\rho} = \frac{3}{4} \sin(\frac{\alpha}{2})$ and
\begin{eqnarray}
C_{i,j,0} &=& {\sqrt{3}\over 4}
          \sin(\frac{\alpha}{2}) (2 v_{i,j,0}-v_{i,j,1} - v_{i-1,j,1})
                                 \nonumber\\
&& - \cos(\frac{\alpha}{4}) s_{i,j-1,1} + 3\cos^3(\frac{\alpha}{4})s_{i,j,0}
\nonumber\\
&&  + (\frac{3}{2}\cos(\frac{\alpha}{4})-\frac{5}{2}\cos^3(\frac{\alpha}{4}))
      (s_{i-1,j,1}+s_{i,j,1})\nonumber\\
&& + \sin(\frac{\alpha}{4})(\frac{5}{2}\cos^2(\frac{\alpha}{4}) -1)
         (u_{i,j,1}-u_{i-1,j,1})\nonumber\\
C_{i,j,1} &=& {\sqrt{3}\over 4}
          \sin(\frac{\alpha}{2}) (v_{i,j,0} + v_{i+1,j,0}- 2 v_{i,j,1})
                 \nonumber\\
&& - \cos(\frac{\alpha}{4}) s_{i,j+1,0}+3\cos^3(\frac{\alpha}{4})s_{i,j,1}
\nonumber\\
&&  + (\frac{3}{2}\cos(\frac{\alpha}{4})-\frac{5}{2}\cos^3(\frac{\alpha}{4}))
      (s_{i+1,j,0}+s_{i,j,0})\nonumber\\
&&  + \sin(\frac{\alpha}{4})(\frac{5}{2}\cos^2(\frac{\alpha}{4}) -1)
      (u_{i+1,j,0}-u_{i,j,0}).
\end{eqnarray}

Taking into account the harmonic potential terms responsible for the central 
{\it i.e} $V_{W}$, non-central, $V_{\Omega}$,  and the bond-bending,  
$V_{C}$ forces,  and using the approximation of the nearest-neighbour 
interaction, we define the phonon Hamiltonian as:
\begin{eqnarray}
H_{ph} &=& \frac12 \sum_{i,j,\rho} \Bigl(
{p_{i,j,\rho}^2\over M}\,+\,
{q_{i,j,\rho}^2\over M}\,+\,
{r_{i,j,\rho}^2\over M}\,+\,k_c\,C_{i,j,\rho}^2\nonumber\\
&&+\, k\,[Wr_{i,j,\rho}^2\,+\,Wl_{i,j,\rho}^2\,+\,Wd_{i,j,\rho}^2
+\,\Omega r_{i,j,\rho}^2\,+\,\Omega l_{i,j,\rho}^2\,
+\,\Omega d_{i,j,\rho}^2 ]\Bigr).
\label{phon-ham}
\end{eqnarray}
Here $p_{i,j,\rho},q_{i,j,\rho}$ and $r_{i,j,\rho}$ are the 
 momenta, canonically conjugate, respectively, to the displacements 
$u_{i,j,\rho},s_{i,j,\rho}$ and $v_{i,j,\rho}$. $M$ is the atom mass, 
$k$ is the  elasticity constant that characterises the central force, and 
$k_c$ is a characteristic constant of the bond-bending force.

Next we assume that each atom in the hexagonal lattice has a well isolated 
non-degenerate electron level ${\cal E}_0$ ($\pi$-orbital in the case of 
a carbon nanotube) and we take into account the hopping interaction, $J$, 
of each atom only with its three nearest neighbours.

The electron-phonon interaction has different sources 
\cite{MDWh,JiDrDr,Kane,WoMah,Mah}. 
Often, the phonon modulation of the hopping interaction  is invoked which,
in the linear approximation with respect to the displacements, gives  
\begin{equation} 
  J_{(i,j,\rho);\delta(i,j,\rho)} = J - G_2 W\delta_{i,j,\rho}, 
\label{J-G2}
\end{equation}
thus leading to the electron phonon interaction proportional to $G_2$.

In general, as has been shown in \cite{WoMah}, in addition to the 
conventional modulation of the hopping interaction one should also take into 
account the interaction between electrons and ions through the Coulomb 
potential which is modulated by the lattice vibrations.

The fields of neighbouring atoms then alter the local electron energy, 
${\cal E}_{i,j,\rho}$, and so, in the same 
linear approximation, we can write 
\begin{equation} 
 {\cal E}_{i,j,\rho} = {\cal 
E}_0 + \chi_1\,(\Wr+\Wl+\Wd) +\chi_2\,C_{i,j,\rho}, 
\label{E_site}
\end{equation}
where $\chi_1$ characterises the influence of the distance  and $\chi_2$ 
 the influence of the bond-bending deviations.

The constants $G_2$, $\chi_1$ and $\chi_2$ determine the strength of the  
electron-phonon coupling. Their calculation would be the subject of 
microscopic considerations. The estimate of $G_2$ for carbon nanotubes was 
obtained in \cite{PStZ}. The coupling, 
through the Coulomb potential, provides an input into all these constants. 
Moreover, the curvature of the lattice  
is an important factor: the  curvature-induced charge polarisation,
present in carbon nanotubes, results in the vibration modulated dipole 
component of the Coulomb potential. As it was indicated in \cite{Kempa}, 
where this fact had been taken into account through the Lydann-Sachs-Teller 
relation, such an electron-phonon interaction is stronger that the one due to 
the deformational potential. In this paper we treat the constants 
$G_2$, $\chi_1$ and 
$\chi_2$ as parameters and we investigate the ground self-trapped state of 
an electron depending on the value of these parameters. The strength of the 
electron-phonon interaction plays a key role in many effects. When the 
coupling constant of this interaction is strong enough it can lead 
to the self-trapping of  a quasiparticle.
By `quasiparticle'  we denote an electron, a hole or an exciton in a nanotube.

The self-trapped states of quasiparticles 
are usually described in the adiabatic approximation which is equivalent to 
the semiclassical consideration in which the vibrational subsystem is 
described as a classical one. Therefore, the state 
of a quasiparticle in such a model is described by the Hamiltonian functional 
\begin{eqnarray}
{\cal H} &=&\,H_{ph}\, +\,\sum_{i,j,\rho} \Bigl({\cal E}_0\,\cmod{\ph}\,
- \frac{1}{2} J\,(\ph^*\phr +  \phr^*\ph\nonumber\\
&+&\ph^*\phl +\phl^*\ph + \ph^*\phd + \phd^*\ph)
\nonumber\\
&+&\chi_1\,\cmod{\ph}\,(\Wr+\Wl+\Wd)\,+\,\chi_2\,\cmod{\ph}\,C_{i,j,\sigma}
\nonumber\\
&+& \frac{1}{2} G_2[(\ph^*\phr+\phr^*\ph)\Wr\,
   +\,(\ph^*\phl+\phl^*\ph)\Wl\,\nonumber\\
&&        +\, (\ph^*\phd+\phd^*\ph)\Wd ]
\Bigr).
\label{EP-hamiltonian}
\end{eqnarray}
Here $H_{ph}$ is given by (\ref{phon-ham}) and $\ph$ is the amplitude of the 
probability of quasiparticle being located on the site $(i,j,\rho)$.

From (\ref{EP-hamiltonian}), after expanding the
$W$'s, $ \Omega$'s and $C$'s, we can derive the following static equations 
for the functions $\ph$, $u$, $v$ and $s$:
\begin{eqnarray}
0 &=&({\cal W }+{\cal E}_0)\pc - J\,\bigl(\pr+\pl+\pd\bigr)
+\chi_1\,\pc\,(W_r+W_l+W_d)+ \chi_2\,\pc\,C\nonumber\\
&+&G_2[\pr W_r +\pl W_l+\pd W_d],
\label{EqnPhi2}
\end{eqnarray}
\begin{eqnarray}
0 &=&k \Bigl[ \sqrt{3} \cos(\frac{\alpha}{4})(W_l-W_r)
             + \cos(\frac{\alpha}{4})(\Omega_l-\Omega_r) + 2 \Omega_d
\Bigr] 
  + k_c \Bigl[ \sin(\frac{\alpha}{4})(\frac{5}{2} \cos^2(\frac{\alpha}{4})-1)
       (C_l -C_r) 
\Bigr] \nonumber\\
&+&\chi_1 (\cmod{\pl}-\cmod{\pr})
        (\frac{\sqrt{3}}{2} \cos(\frac{\alpha}{4}))
 +\chi_2 (\cmod{\pl}-\cmod{\pr})
           \sin(\frac{\alpha}{4})(\frac{5}{2} \cos^2(\frac{\alpha}{4})-1)
\nonumber\\
&+& {\sqrt{3}\over 2} \cos(\frac{\alpha}{4}) G_2 
   (\pc^*\pl+\pl^*\pc-\pc^*\pr+\pr^*\pc),
\label{Eqnu2}
\end{eqnarray}

\begin{eqnarray}
0 &=&k \Bigl[ (2 W_d -W_r-W_l) +\sqrt{3}(\Omega_r+\Omega_l)
\Bigr] 
  + k_c \Bigl[ \frac{\sqrt{3}}{4}\sin(\frac{\alpha}{2})
        ( 2 C + C_r + C_l)
\Bigr] \nonumber\\
&+&\chi_1 \frac{1}{2} (2\cmod{\pd}-\cmod{\pr}-\cmod{\pl})
+\chi_2 (2\cmod{\pc}+\cmod{\pr}+\cmod{\pl})
        (\frac{\sqrt{3}}{4}\sin(\frac{\alpha}{2}) )
\nonumber\\
&+& \frac{1}{2} G_2 
   (-\pc^*\pr-\pr^*\pc-\pc^*\pl-\pl^*\pc + 2\pc^*\pd+2\pd^*\pc ),
\label{Eqnv2}
\end{eqnarray}

\begin{eqnarray}
0 &=&k \Bigl[ \sqrt{3} \sin(\frac{\alpha}{4}) (Wr+Wl )
             +\sin(\frac{\alpha}{4}) (\Omega_r +\Omega_l )
       \Bigr] \nonumber\\
  &+& k_c \Bigl[
(\frac{3}{2}\cos(\frac{\alpha}{4})-\frac{5}{2}\cos^3(\frac{\alpha}{4}))
   (C_r+C_l) 
 -\cos(\frac{\alpha}{4})C_d + 3\cos^3(\frac{\alpha}{4})C
\Bigr] \nonumber\\
&+&\chi_1 \frac{\sqrt{3}}{2} \sin(\frac{\alpha}{4}) 
     (2\cmod{\pc}+\cmod{\pr}+\cmod{\pl})\nonumber\\
&+&\chi_2 \Big(
(\frac{3}{2}\cos(\frac{\alpha}{4})-\frac{5}{2}\cos^3(\frac{\alpha}{4}))
   (\cmod{\pr}+\cmod{\pl}) 
-\cos(\frac{\alpha}{4})\cmod{\pd}
   + 3\cos^3(\frac{\alpha}{4})\cmod{\pc}
   \Bigr)\nonumber\\
&+& G_2 {\sqrt{3}\over 2} \sin(\frac{\alpha}{4})
     (\pc^*\pr+\pr^*\pc+\pc^*\pl+\pl^*\pc).
\label{Eqns2}
\end{eqnarray}
Here we have used the notation $\pc, \pr, \pl, \pd $ for the functions 
on the lattice site $\{i,j,\rho \}$ and on its corresponding 3 neighbours.

\section{Numerical results}

In this section we  describe various numerical stationary solutions of our 
equations. Note that our equations possess several parameters: 
$J$, $k$, $k_c$, $G_2$, $\chi_i$ and the lattice spacing $d$.
For the numerical calculations  we need dimensionless quantities
and so we perform the following change of variables:
\begin{eqnarray}
&& u=lU; \qquad  v=lV; \quad 
s=lS;\quad k={\kappa\over l^2}J;\qquad k_c={\kappa_c\over l^2}J;\nonumber\\
&&  \chi_1={X_1\over l^2}J; \qquad
\chi_2={X_2\over l^2}J; \qquad G_2={G\over l^2}J; \qquad
{\cal E}_0=\tilde{\cal E}_0 J.
\end{eqnarray}
First we choose to measure the energy in units of $J$ and so we put $J=1$. 
Then ${\cal E}_0$ is dimensionless. Next we choose the unit of 
length $l$ to be such that $\kappa=1$. From the expressions above we see that 
this corresponds to $l=\sqrt{J/k}$. For the carbon nanotube the parameters 
take the following values  
$J\approx 2.4 eV $ \cite{SaiDrDr,Mah,Dai}, $d=1.42\times 10^{-10} m$ 
\cite{SaiDrDr}, $k=36.5 \times 10^4 {dyn/ cm}$ \cite{SaiDrDr}. 
Using these values, we get
 $l=(3.24-3.7)\times 10^{-11} m$. Thus, the length scale we use is smaller 
than the lattice length $d\approx 4\,l$.
For carbon nanotubes the dimensionless electron-phonon coupling constant 
$G={G_2 l/ J} \sim 0.6-0.74$.

We have derived numerical solutions of our model for all values of $n$ 
between  $3$ and $20$ for several ranges of other parameters. In all our 
calculations we have taken $k=1$, $k_c=0.2$. Then fixing
the values of two of the three parameters $X_1$, $X_2$ and $G$ we have 
varied the value of the 3rd one between $0$ and $5$, and $\chi_1$ and 
$\chi_2$ mostly between 0 and 1. We have also set the lattice 
parameter $a=1$ and, in our discussions, we will refer to the main axis of the 
nanotube as $y$ while $x$  will run along the circumferences of the nanotube.

Of course, we have used periodic boundary conditions along the $x$ axis, while 
for the ends of the nanotube we have used free boundary conditions. This 
allows not only for
a global translation of the nanotube but also for a dilation or contraction of 
the lattice along its axis. To exclude global 
translation of the lattice,  we have pinned a single lattice site, at the 
bottom of the lattice, using a quadratic potential. Using such a potential, 
rather than fixing the position to a fixed value, is smoother from the 
numerical point of view.

\subsection{Numerical solutions}

\begin{figure}[htbp]
\unitlength1cm \hfil
\begin{picture}(16,8)
 \epsfxsize=8cm 
  \epsffile{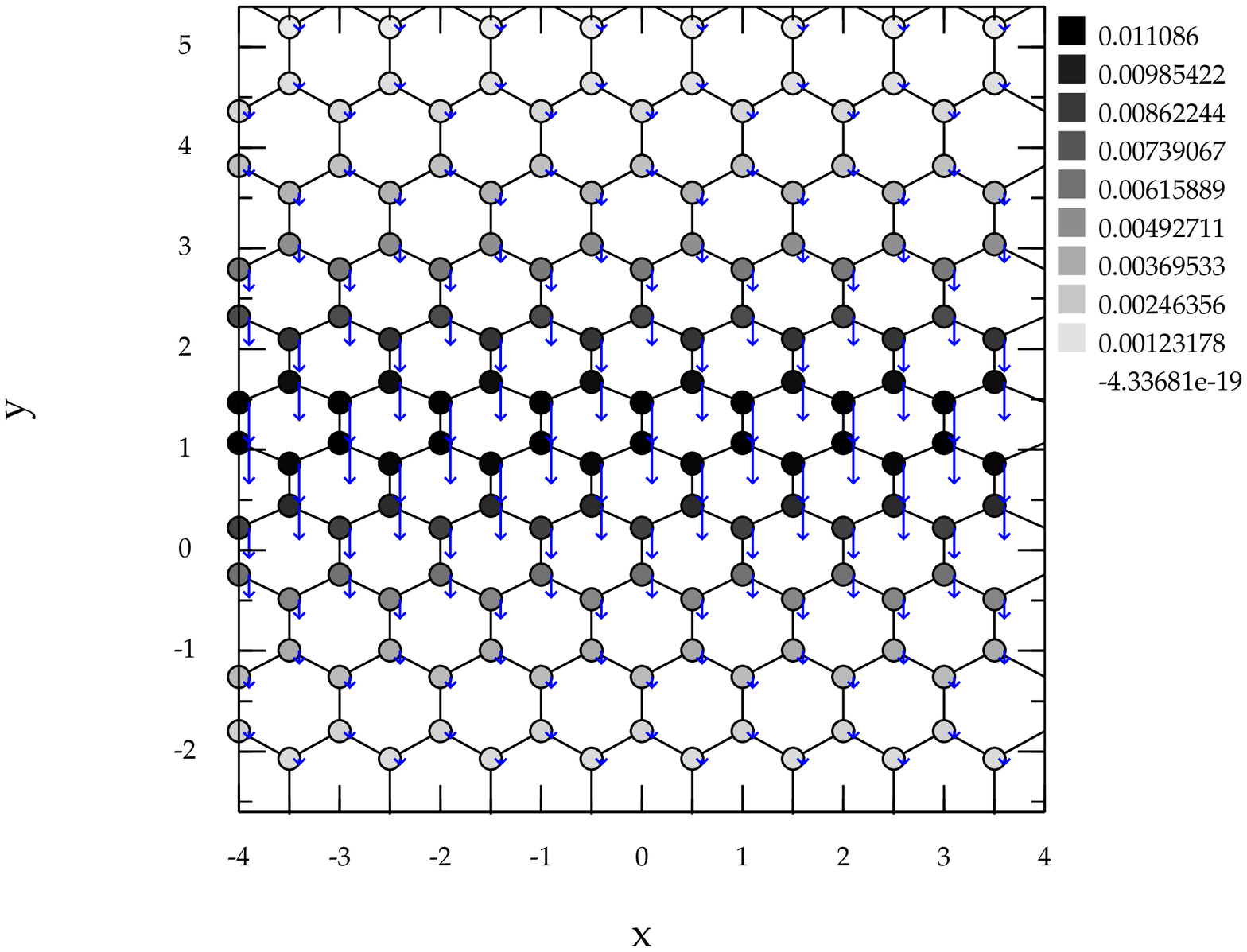}
 \epsfxsize=8cm 
  \epsffile{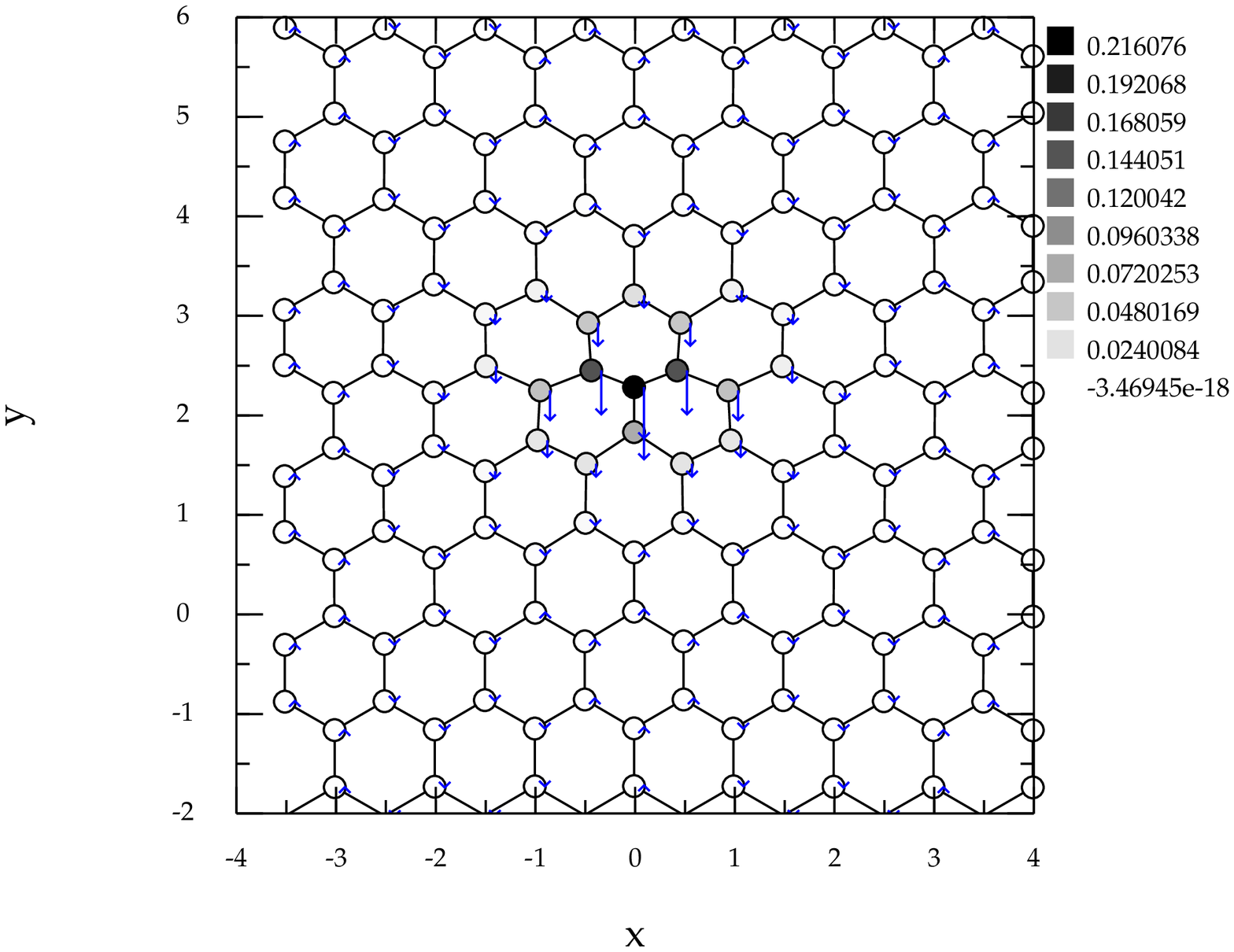}
\end{picture}
\caption{\label{Sol1}
Solution for $n=8$, $X_1=0.6$, $X_2=0.2$ and a) $G=1$. b)  $G=1.5$.
The deformed lattice is plotted with small circles at the vertices 
representing the electron density (see the grey scale on the right; 
white corresponding to $|\varphi|^2 = 0$).
The arrows at each vertex represent the $s$ field; a down/up arrow 
represents a displacement towards/away from the centre of the 
tube. The displacements are plotted in dimensionless units for $G=1.5$ and 
amplified by a factor of $10$ for $G=1$. 
}
\end{figure}

\begin{figure}[htbp]
\unitlength1cm \hfil
\begin{picture}(16,8)
 \epsfxsize=8cm 
   \epsffile{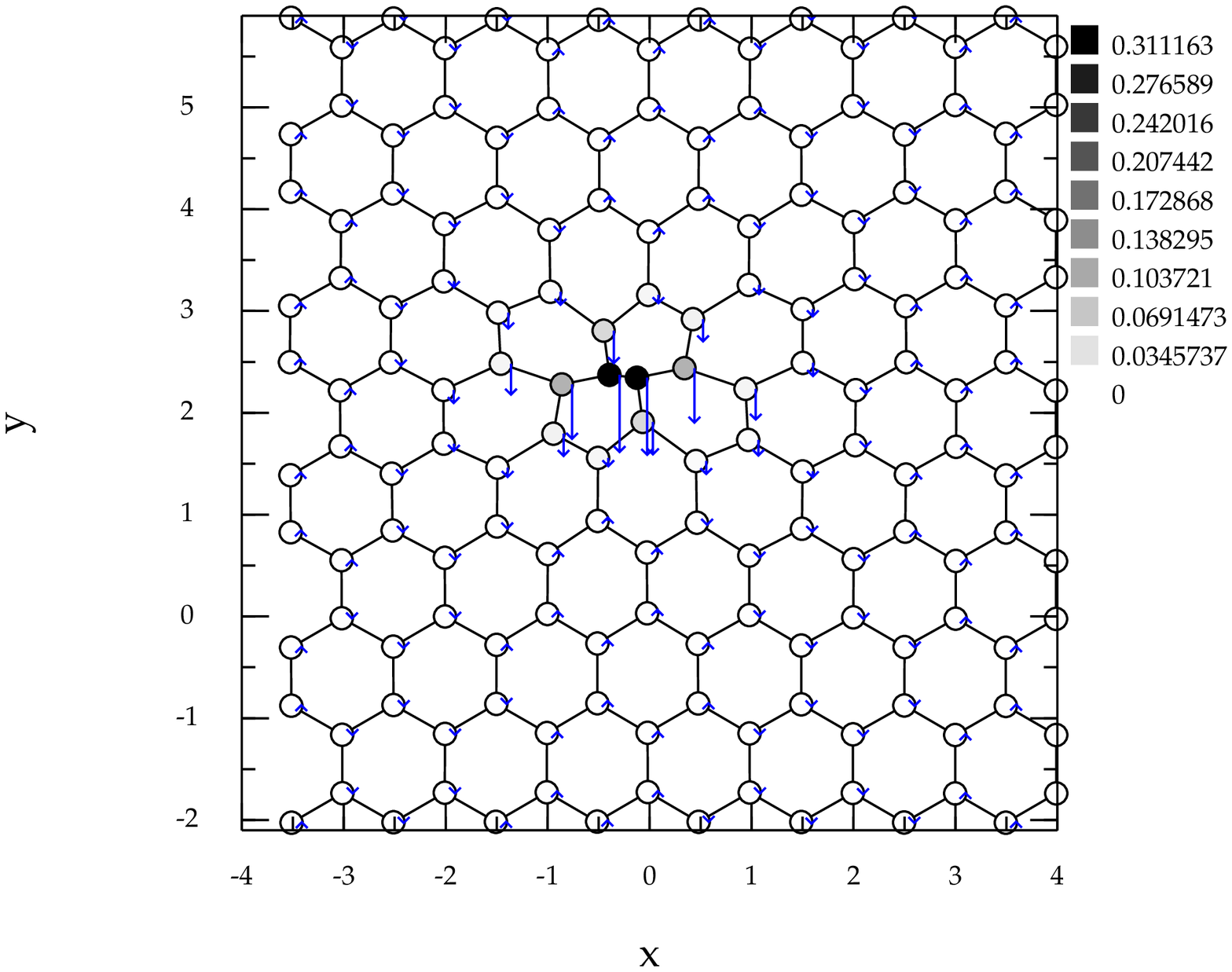}
 \epsfxsize=8cm 
   \epsffile{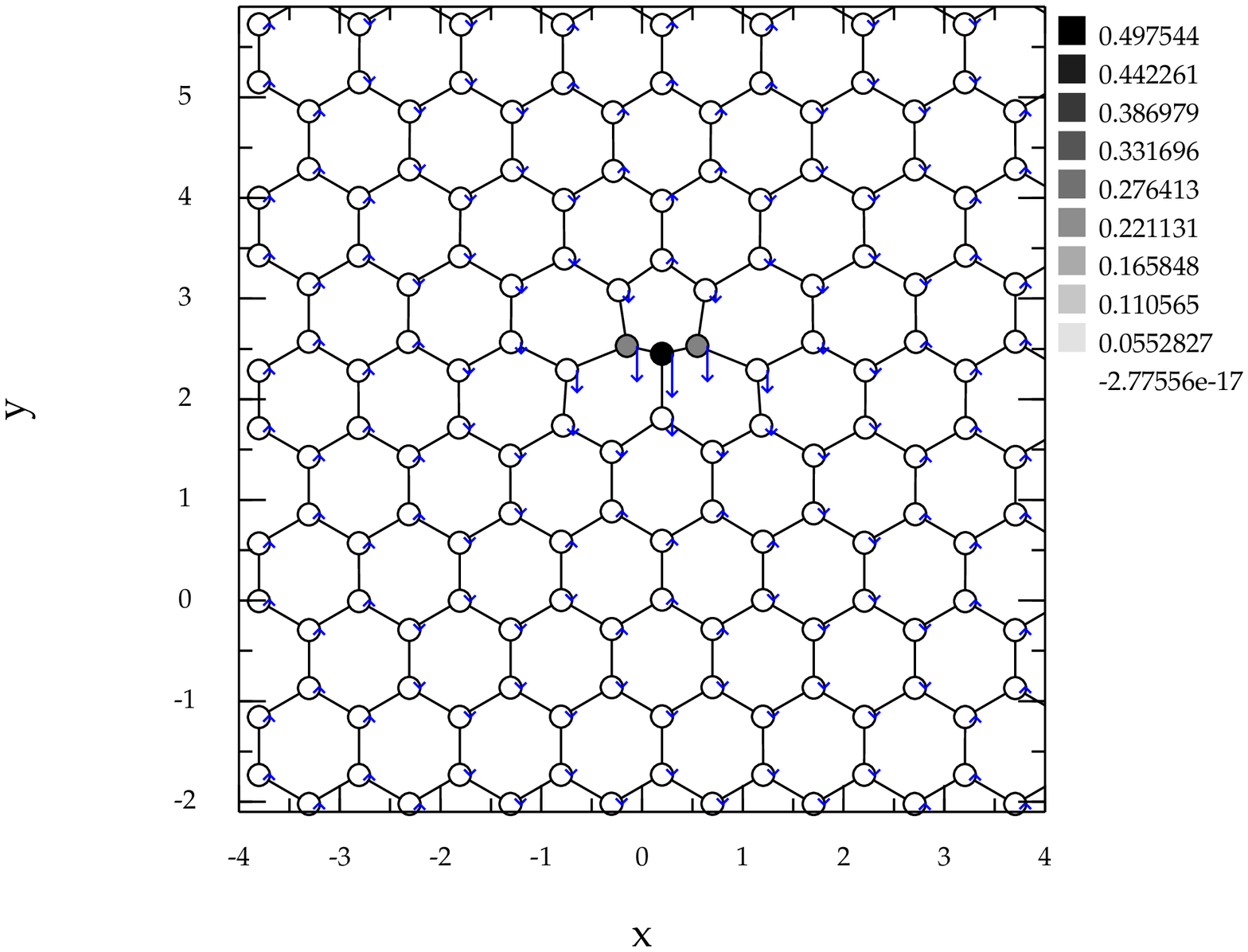}
\end{picture}
\caption{\label{Sol2}
Solution for $n=8$, $X_1=0.6$, $X_2=0.2$ and a) $G=2$ b)  $G=5$.
The deformed lattice is plotted with small circles at the vertices 
representing the electron density (see the grey scale on the right; 
white corresponding to $|\varphi|^2 = 0$).
The arrows at each vertex represent the $s$ field; a down/up arrow represents 
a displacement towards/away from the centre of the 
tube. The displacements are plotted in dimensionless units for $G=2$ and in 
physical units (dimensionless units divided by $4$) for $G=5$.  
}
\end{figure}

\begin{figure}[htbp]
\unitlength1cm \hfil
\begin{picture}(16,8)
 \epsfxsize=8cm \epsffile{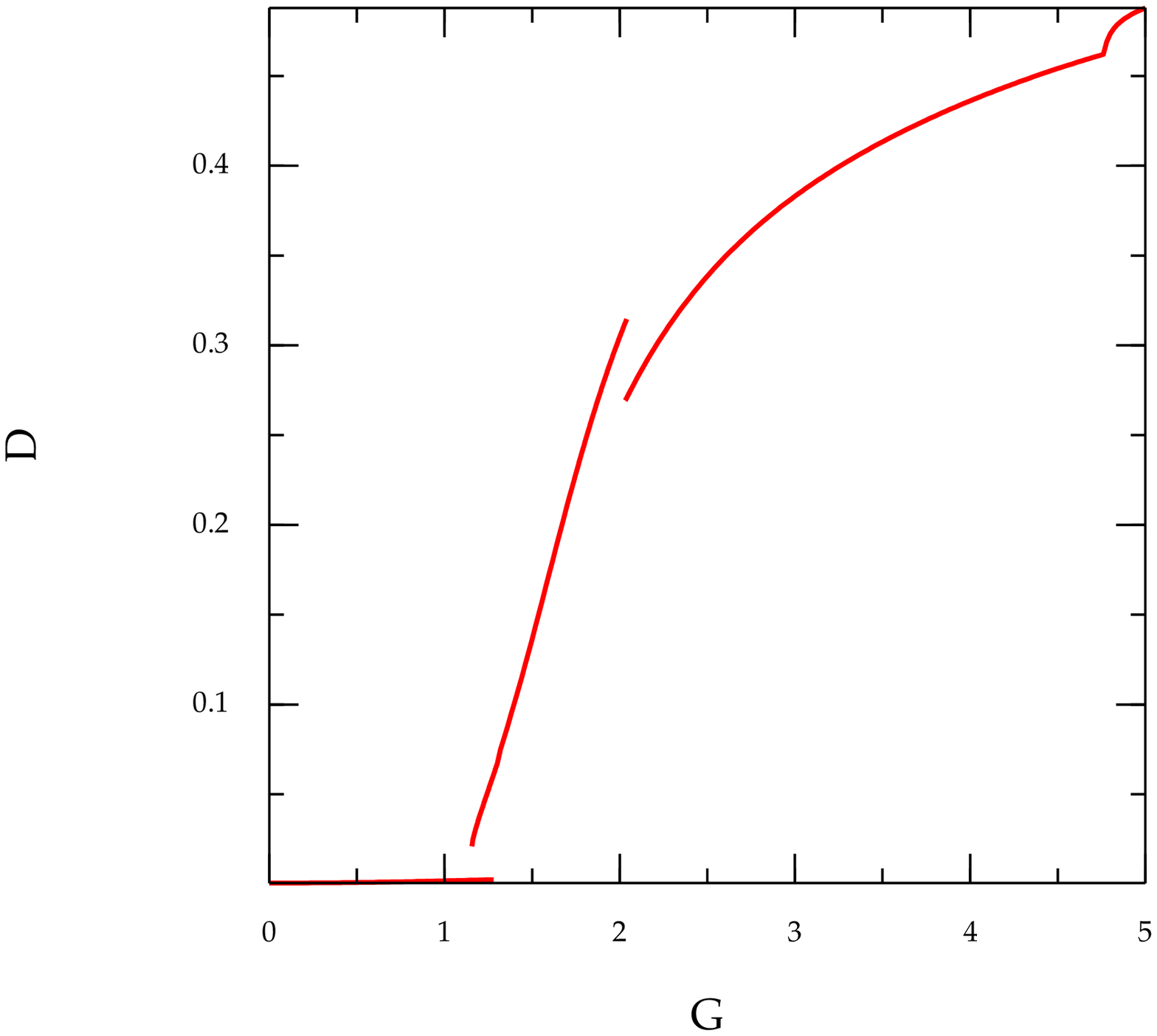}
 \epsfxsize=8cm \epsffile{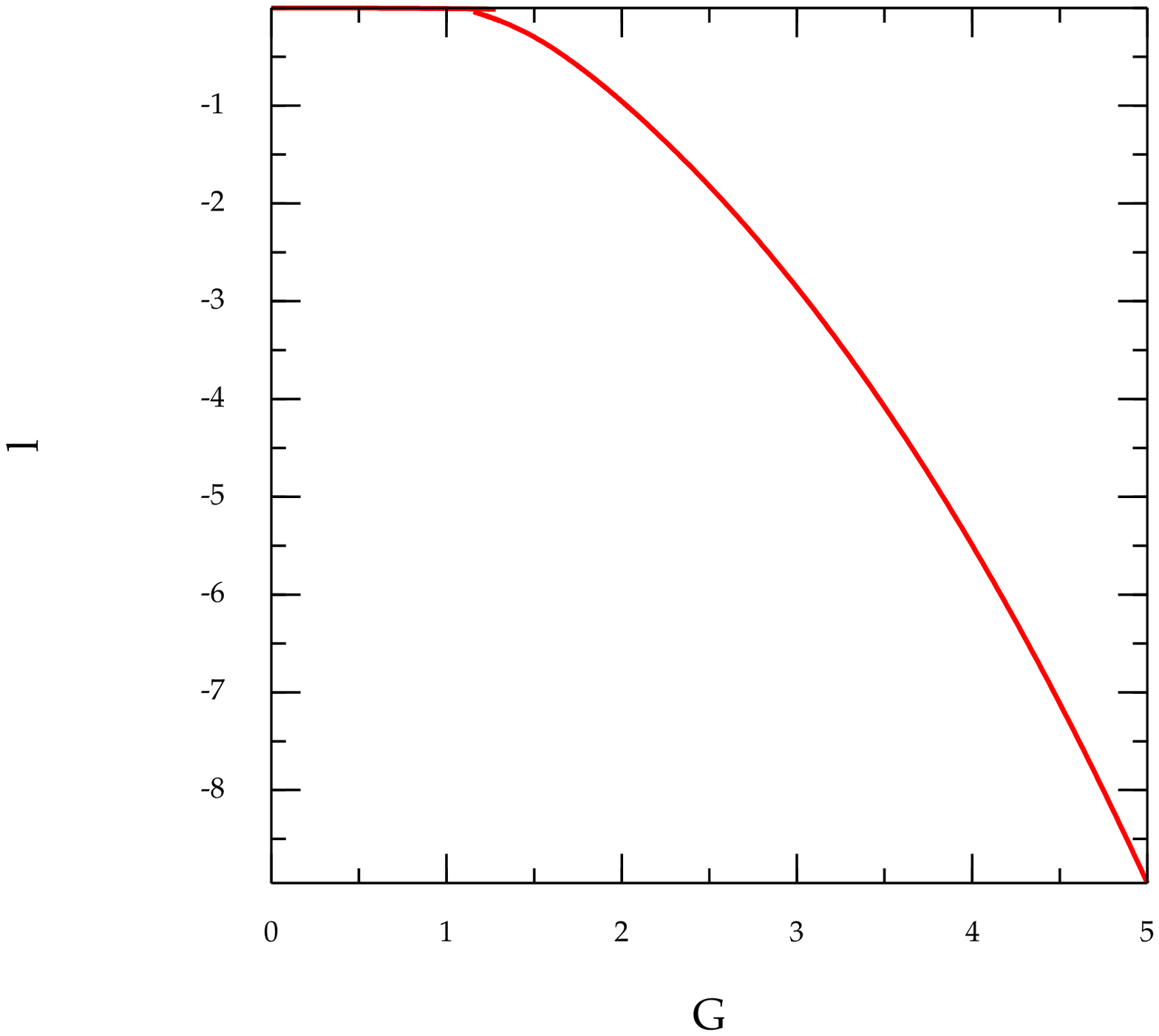}
\end{picture}
\caption{\label{Gan20}
The maximum of probability, $D = \hbox{Max}_{(i,j,\rho)}\,|\varphi_{i,j,\rho}|^2$ 
and the eigenvalue 
$\Lambda$ for $X_1=0.6$, $X_2=0.2$ as a function of $G$ for $n=20$.}
\end{figure}

\begin{figure}[htbp]
\unitlength1cm \hfil
\begin{picture}(16,8)
 \epsfxsize=8cm \epsffile{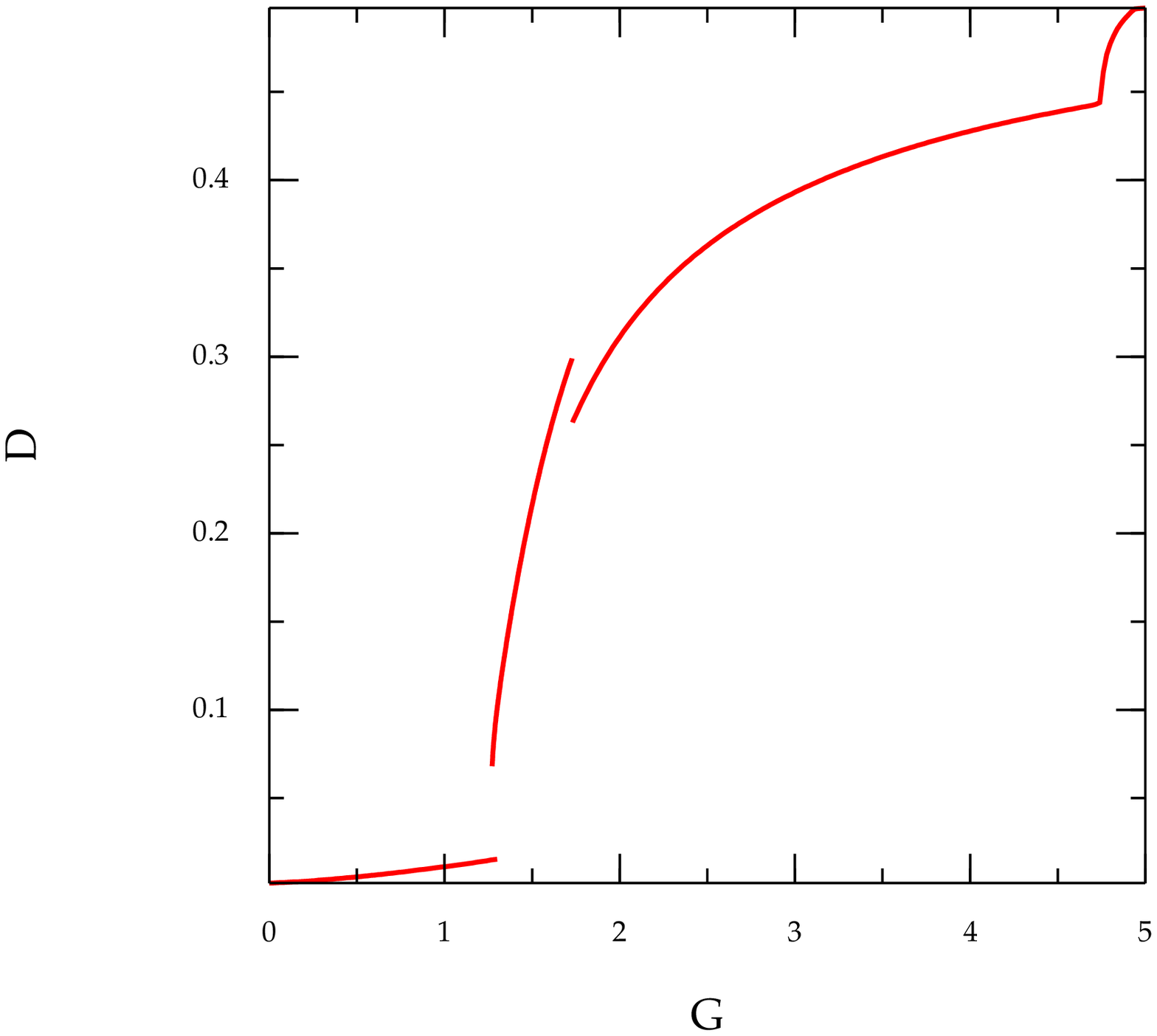}
 \epsfxsize=8cm \epsffile{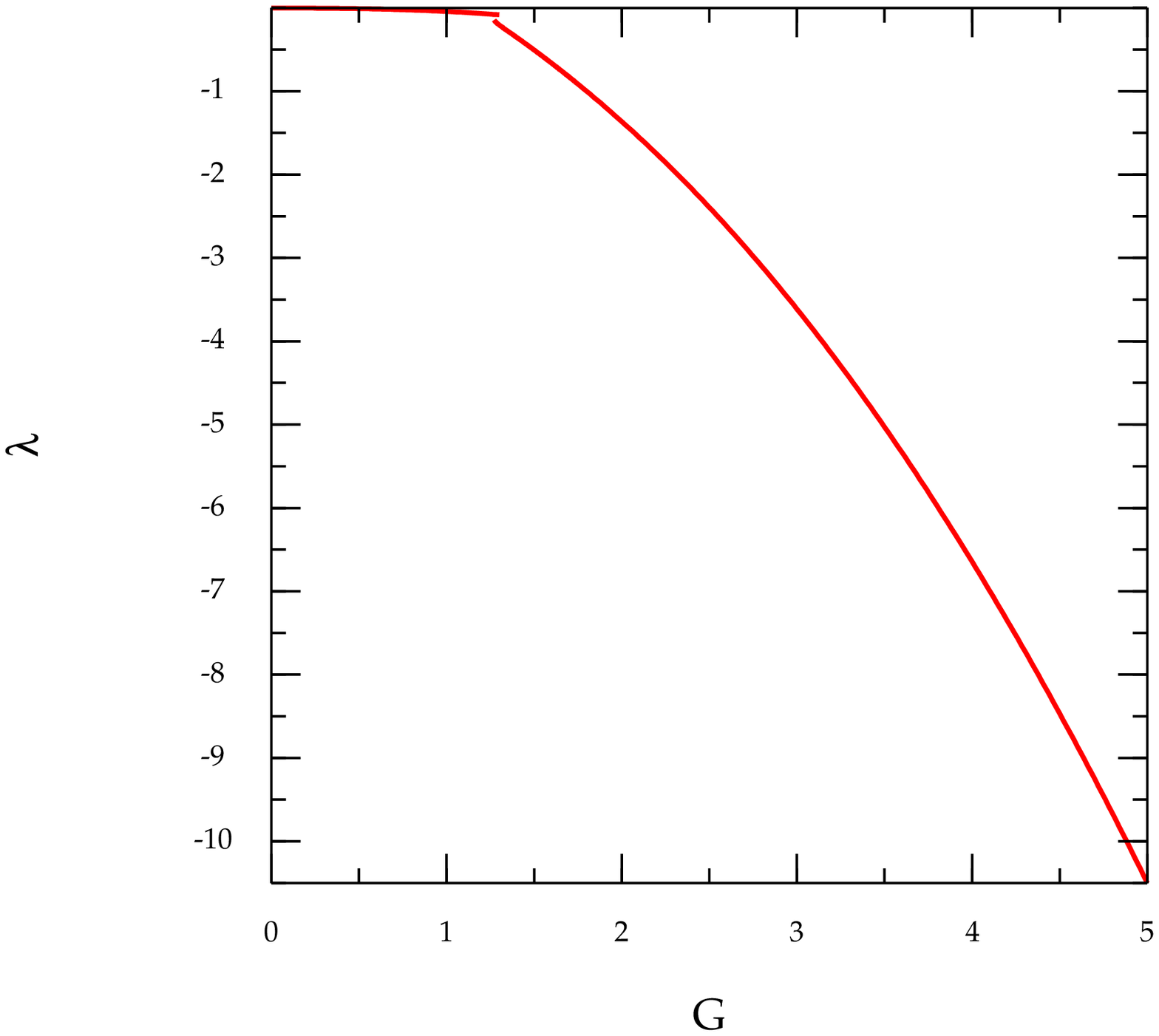}
\end{picture}
\caption{\label{Gan8}
The maximum of probability, $D = \hbox{Max}_{(i,j,\rho)}\,|\varphi_{i,j,\rho}|^2$ 
and the eigenvalue $\Lambda$  for $X_1=0.6$, $X_2=0.2$ as a function of 
$G$ for $n=8$.}
\end{figure}

\begin{figure}[htbp]
\unitlength1cm \hfil
\begin{picture}(16,8)
 \epsfxsize=8cm \epsffile{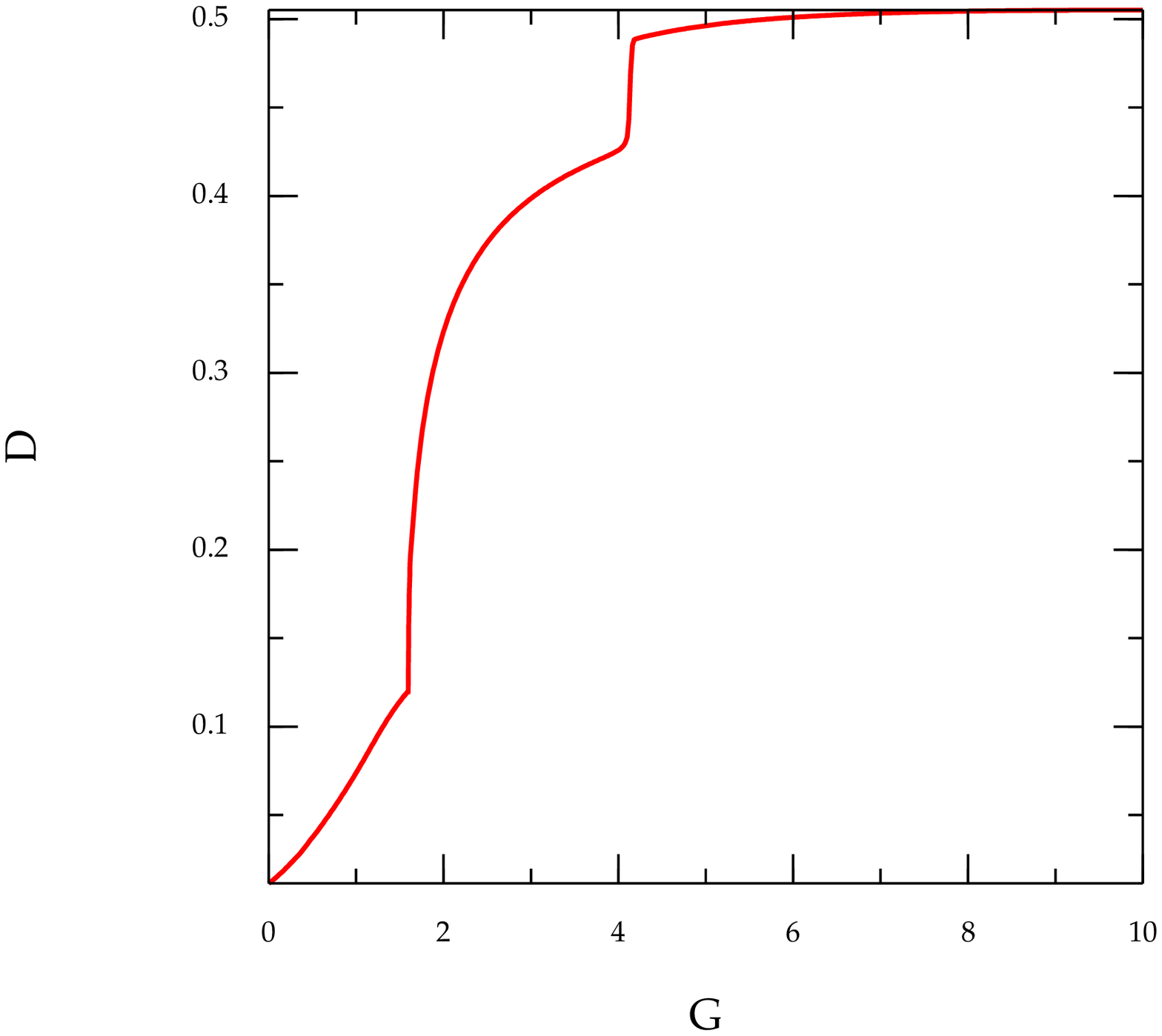}
 \epsfxsize=8cm \epsffile{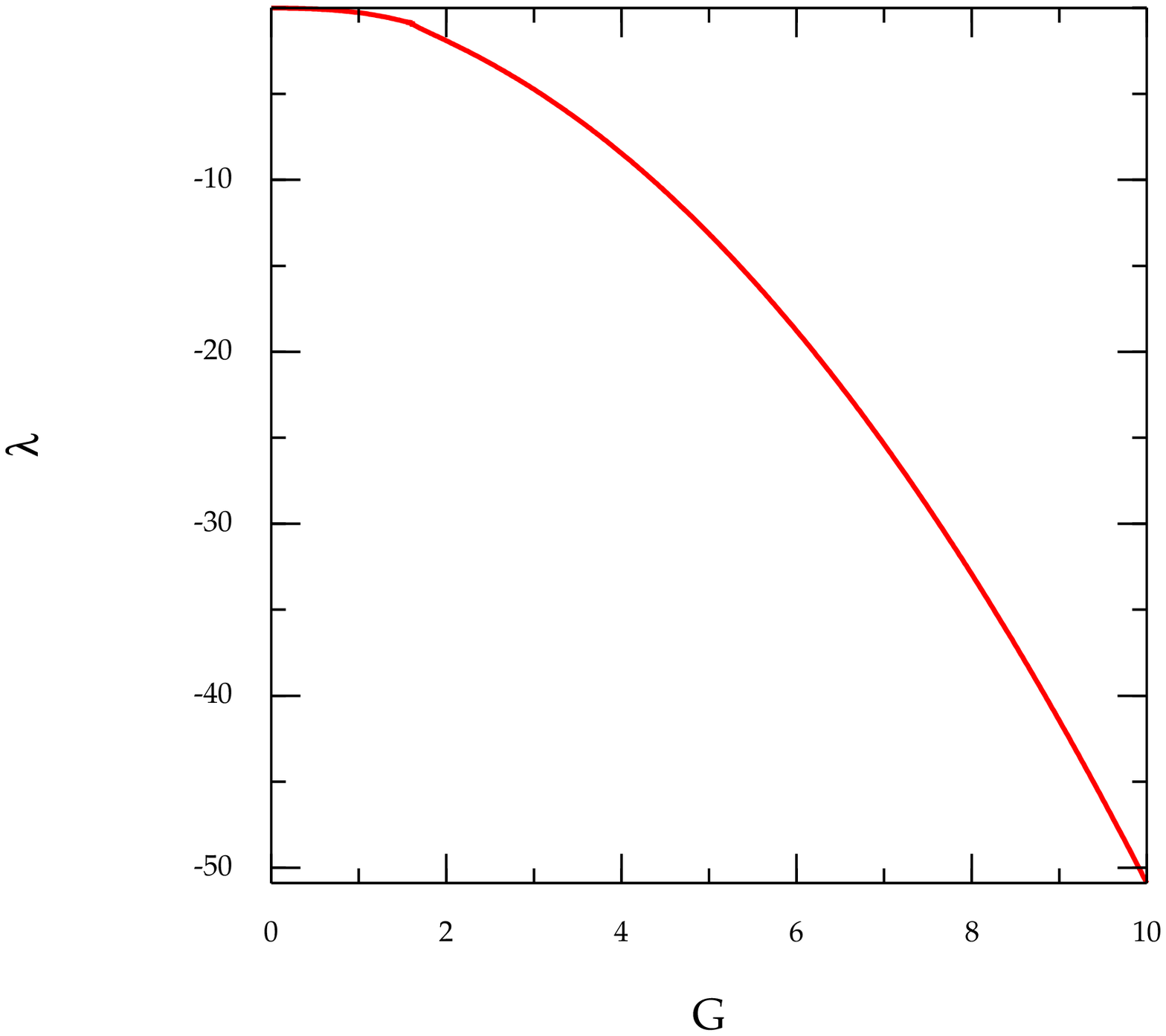}
\end{picture}
\caption{\label{Gan3}
The maximum of probability, $D = \hbox{Max}_{(i,j,\rho)}|\varphi|^2$ and the 
eigenvalue $\Lambda$ for $X_1=0.6$, $X_2=0.2$ as a function of $G$ for $n=3$.}
\end{figure}

\begin{figure}[htbp]
\unitlength1cm \hfil
\begin{picture}(8,8)
 \epsfxsize=8cm \epsffile{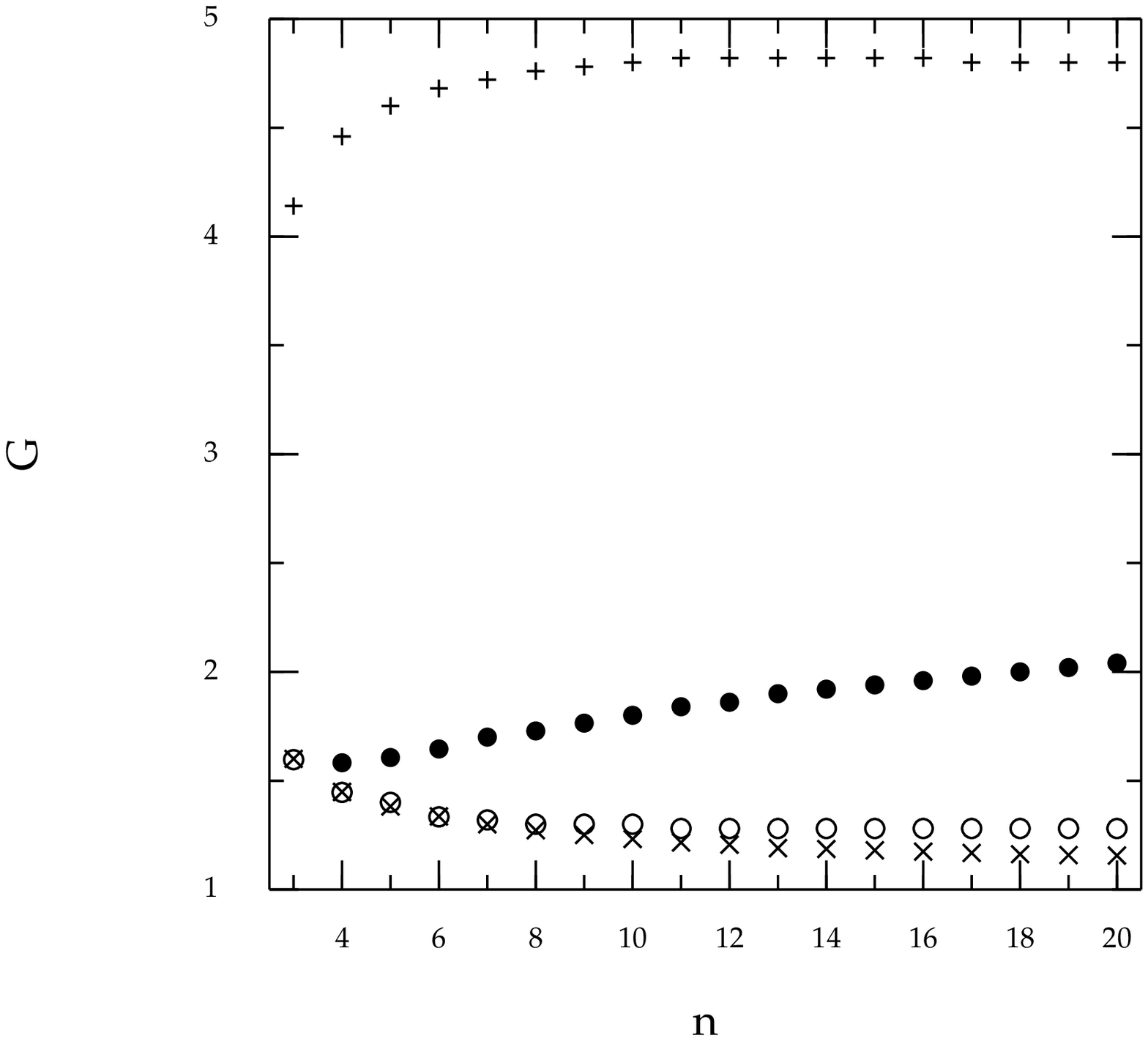}
\end{picture}
\caption{\label{TransG2}
Domain of solutions as a function of $n$ for solutions with
$X_1=0.6$, $X_2=0.2$. 
$\circ$: upper value of $G$ for ring like solution (type I). 
$\times$: lower value of $G$ for solutions of type II.
$\bullet$ : Value of $G$ at the transition between solutions of type II and 
type III.
$+$ : Value of $G$ at the transition between solutions of type III and 
type IV.}
\end{figure}

\begin{figure}[htbp]
\unitlength1cm \hfil
\begin{picture}(16,8)
 \epsfxsize=8cm \epsffile{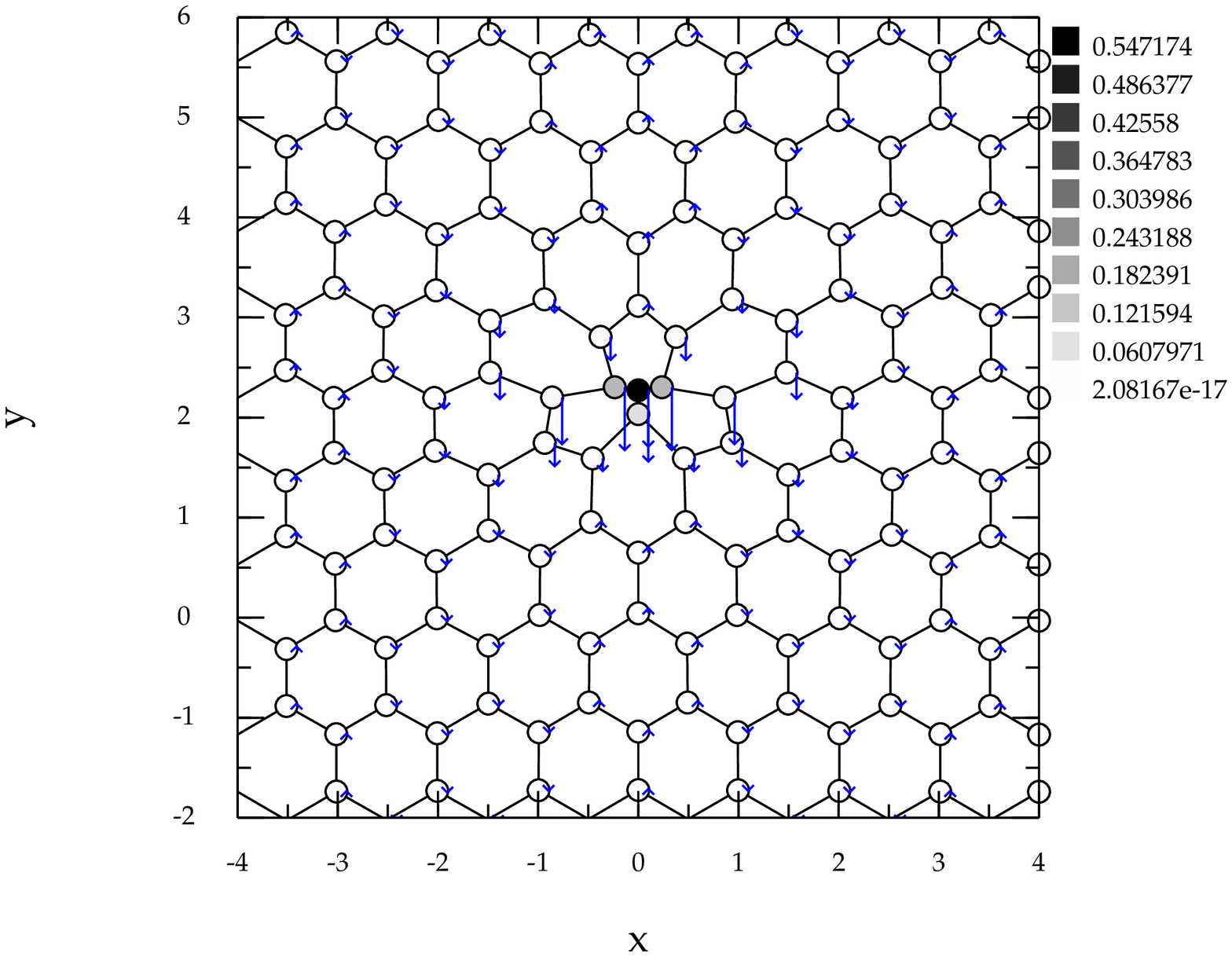}
 \epsfxsize=8cm 
    \epsffile{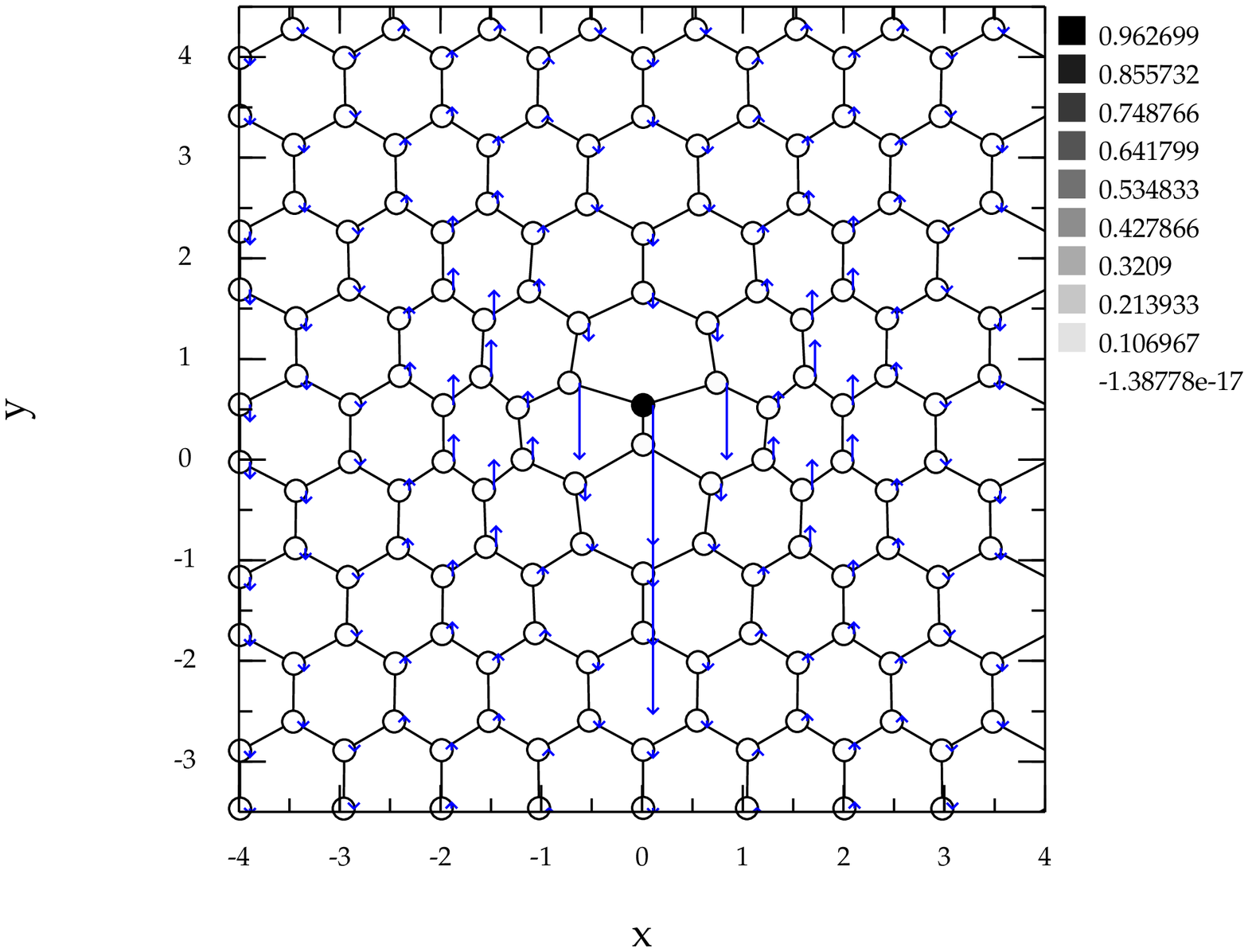}
\end{picture}
\caption{\label{Sol3}
Solution for $n=8$, $G=0.6$ and a) $X1=2$, $X_2=0$ 
b) $X1=0.6$, $X_2=1.5$.  }
The deformed lattice is plotted with small circles at the vertices 
representing the electron density (see the grey scale on the right; 
white corresponding to $|\varphi|^2 = 0$).
The arrows at each vertex represent the $s$ field; a down/up arrow represents 
a displacement towards/away from the centre of the 
tube. The displacements are plotted in dimensionless units. 
\end{figure}

We have found that, depending on the values of the parameters, the system 
of equations (\ref{EqnPhi2}-\ref{Eqns2}) possesses several types of solutions. 
Some of them, sometimes, coexist at the
same values of the coupling constant. Some of these solutions are shown in 
Figs. \ref{Sol1}-\ref{Sol3} where we have plotted the 
hexagonal lattice unfolded onto a plane. The periodic boundary conditions imply
that the left and right hand sides of the lattice are connected together (the
lines joining these sites are not drawn in our figures, they are implicit).
The position of each site is given 
by its equilibrium value translated by the deformations 
$u$ and $v$ at that lattice site. 
The $s$ field, on the other hand, is represented by an arrow, on the
same scale, which points downwards if the $s$ field points towards the centre 
of the tube, {\it i.e.} if the tube is pinched,  or upwards if the field
points away from the tube, {\it i.e.} if the tube is dilated.

Most of the solutions we present here correspond to the $n=8$ case but the 
solutions for other 
values of $n$ look very much the same. Occasionally we will add some comments
on this dependence. In the figures we only present a small section of the
nanotube we have taken; for clarity of the picture we do show the full width 
of the tube, along its periodic direction, but we restrict ourselves to the 
section of the tube 
along its axis where the soliton is located. The solutions were actually 
computed on a grid ranging from $-50$ to $+50$.

In our plots we have used the dimensionless units described above, which 
correspond to displacements 4 times larger than the real displacements of the 
nanotube, 
except for Figure \ref{Sol1}.a where we had to amplify the dimensionless 
units by a factor of 10 and Figure \ref{Sol2}.b where we have used the 
physical units because the displacements in this case were already 
sufficiently large.
In the figures, each lattice site is represented by a circle, the inside of 
which, is painted in 
 shades of black describing the value of the polaron probability density 
$|\varphi|^2$. A white inside of the circle corresponds to a very small 
density,  
while black is used for the largest density. The actual shade scale is 
given on the right hand side of each figure. Note that the axis of the 
nanotube runs vertically in our figures. 

When the electron-phonon coupling constants are small the solutions are 
invariably 1D bell-shape solitons wrapped around the 
nanotube.  This is seen in Fig. \ref{Sol1}.a where the amplitudes of 
the displacements have been amplified $40$ times compared to their physical 
values. 
In what follows, we name these solutions - type I solutions.
We see that for type I solutions the polaron probability distribution 
looks like a 1D Davydov soliton along the axis of the nanotube 
which is spread uniformly around it.
 For our choice of parameters the nanotube is contracted
at the position of the soliton and the tube itself is also pinched. 
Note that if we choose the 3 electron-phonon coupling constants with the 
opposite signs, the displacements fields $u$, $v$ and $s$ will change signs 
and the nanotube will become extended and inflated.

Note also that the soliton drags the whole nanotube towards itself. In other 
words, the contraction of the lattice at the soliton position is not 
compensated by an appropriate dilation elsewhere.

When we increase the electron-phonon coupling constants above some critical 
value, the radially symmetrical solutions do not correspond to the minimum 
value of energy and the polaron localises on a few sites.
We show some typical solutions in Figs. \ref{Sol1}.b, \ref{Sol2} and 
\ref{Sol3}. 
The solution at $X_1=0.6$, $X_2=0.2$, $G=1.5$  is shown in Fig. \ref{Sol1}.b):
 the polaron probability density reaches a 
maximum on a single lattice site, but the polaron is spread over several 
lattice sites around the position of its maximum density. The lattice is 
slightly contracted but the lattice sites, where the polaron is localised, are
slightly distorted into the nanotube. 
We name such solutions - 'the type II solutions'.

In Fig. \ref{Sol2}.a, we present a solution
which is also localised on a few sites but for which the maximum of 
the polaron density is evenly distributed between 2 lattice sites. This 
solution 
corresponds to the same values of parameters as figure \ref{Sol1}.b 
except that $G=2$. In this case, the lattice is also contracted, the polaron 
is more localised, and the lattice sites where the polaron is 
localised are also moved into the nanotube. Notice that the two sites at 
which 
the polaron density reaches its maximum  have moved to be very close to each 
other. We name such solutions - 'the type III solutions'.

In Fig. \ref{Sol2}.c, we present a solution corresponding to the same 
values of the parameters as figures \ref{Sol1}.b and \ref{Sol2}.a but for which
$G=5$. This time the polaron is  localised mainly on 3 lattice sites: it has 
a probability density of about $0.5$  on one site and
of $0.25$ on either of the two neighbouring sites. The lattice 
deformation in this case is  very large and so we had to scale it down to the
physical scale. Once again, the lattice is contracted at the position of the
polaron and the lattice sites where the polaron is located have all moved 
into the nanotube.
We name these solutions - 'the type IV solutions'.

The four classes of solutions we have presented so far, correspond to the same 
family of
solutions obtained when one varies $G$ and takes $X_1=0.6$ and $X_2=0.2$. 
In Figs. \ref{Gan20}-\ref{Gan3} we present the maxima of the polaron 
probability densities ($D = \hbox{Max}_{(i,j,\rho)}\,|\varphi_{i,j,\rho}|^2$) 
and the eigenvalues $\Lambda$ of these solutions
as a function of $G$. We present the data for the cases of $n=20,$ $8$ and $3$.
The solutions shown in figures \ref{Sol1} and  \ref{Sol2} correspond to
the 4 types of configurations presented  in figures \ref{Gan8}. 
In particular we note that the transitions between the different types of 
solutions are always sharp and that sometimes there is an overlap: 
two different types of 
solutions coexist for the same values of the parameters. 

We also note that the diameter of the nanotube does not affect very much the 
types of solutions that exist, except for the location of the transition 
between the different types of solutions. On the other hand, the binding 
energies of the solutions are larger for smaller values of $n$.

In Fig. \ref{TransG2} we present the values of $G$ for which the various 
types of solutions exist as a function of $n$.
We see that there is an overlap between solutions of type I and II and that 
this overlap is larger for larger values of $n$.


In Fig. \ref{Sol3}.a, $X_1=2$, $X_2=0$, $G=0.6$ we present a solution 
which appears to be rather different from the previous ones. This time, the 
polaron is localised strongly on 4 sites, but with a density of about 
$0.55$ at the central site.
Note that the deformation of the lattice is nearly symmetric this time.
A slight asymmetry is induced by the bending of the nanotube.
We name these solutions - 'the type V solutions'.

Finally in Fig. \ref{Sol3}.b we present a solution that is strongly 
localised on a single site. Note that this time the lattice is very strongly 
deformed. We name such solutions - 'the type VI solutions'.

\subsection{Solutions for $X_1=0.6$, $G=0.6$ and varying $X_2$}

\begin{figure}[htbp]
\unitlength1cm \hfil
\begin{picture}(16,8)
 \epsfxsize=8cm \epsffile{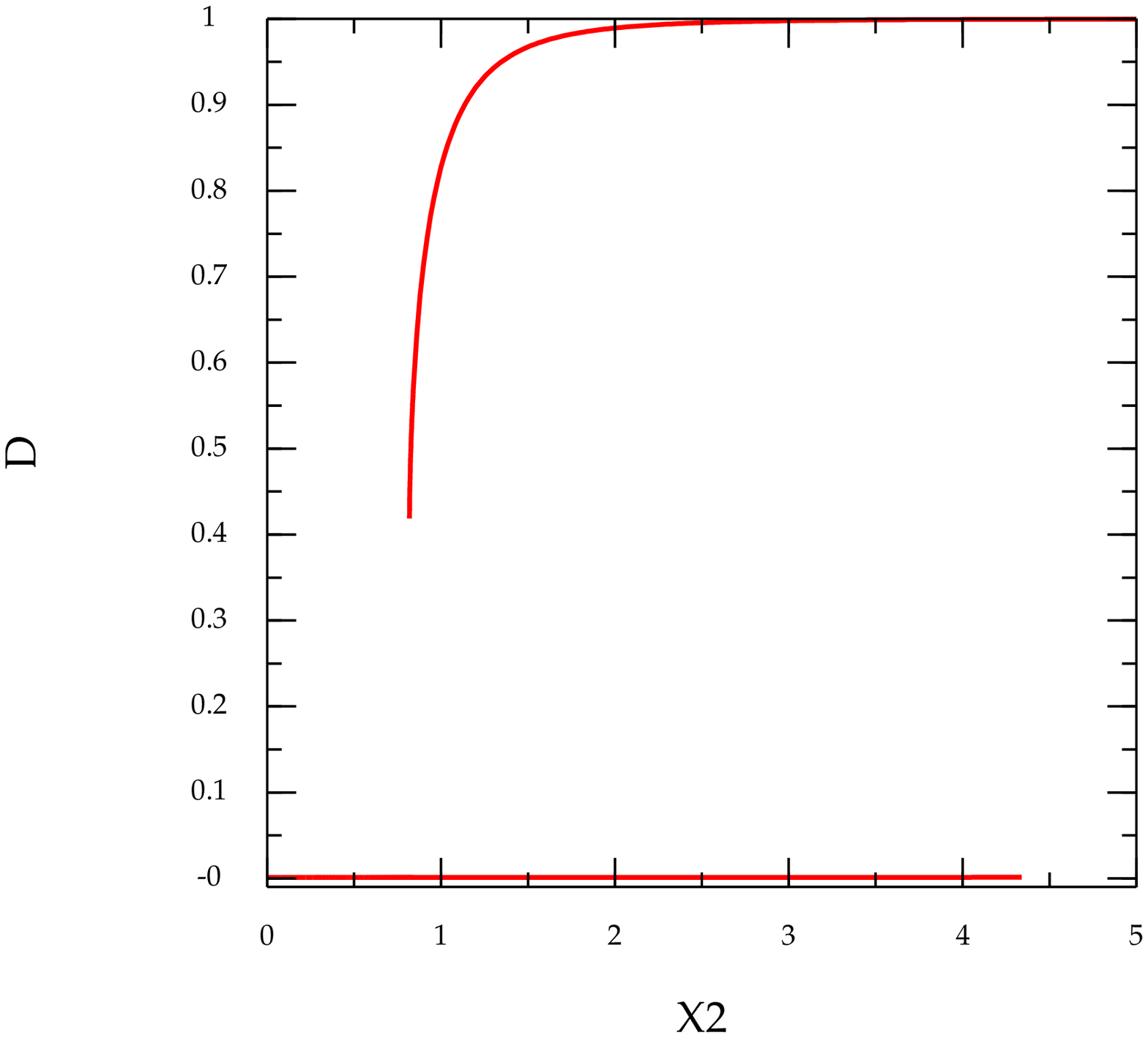}
 \epsfxsize=8cm \epsffile{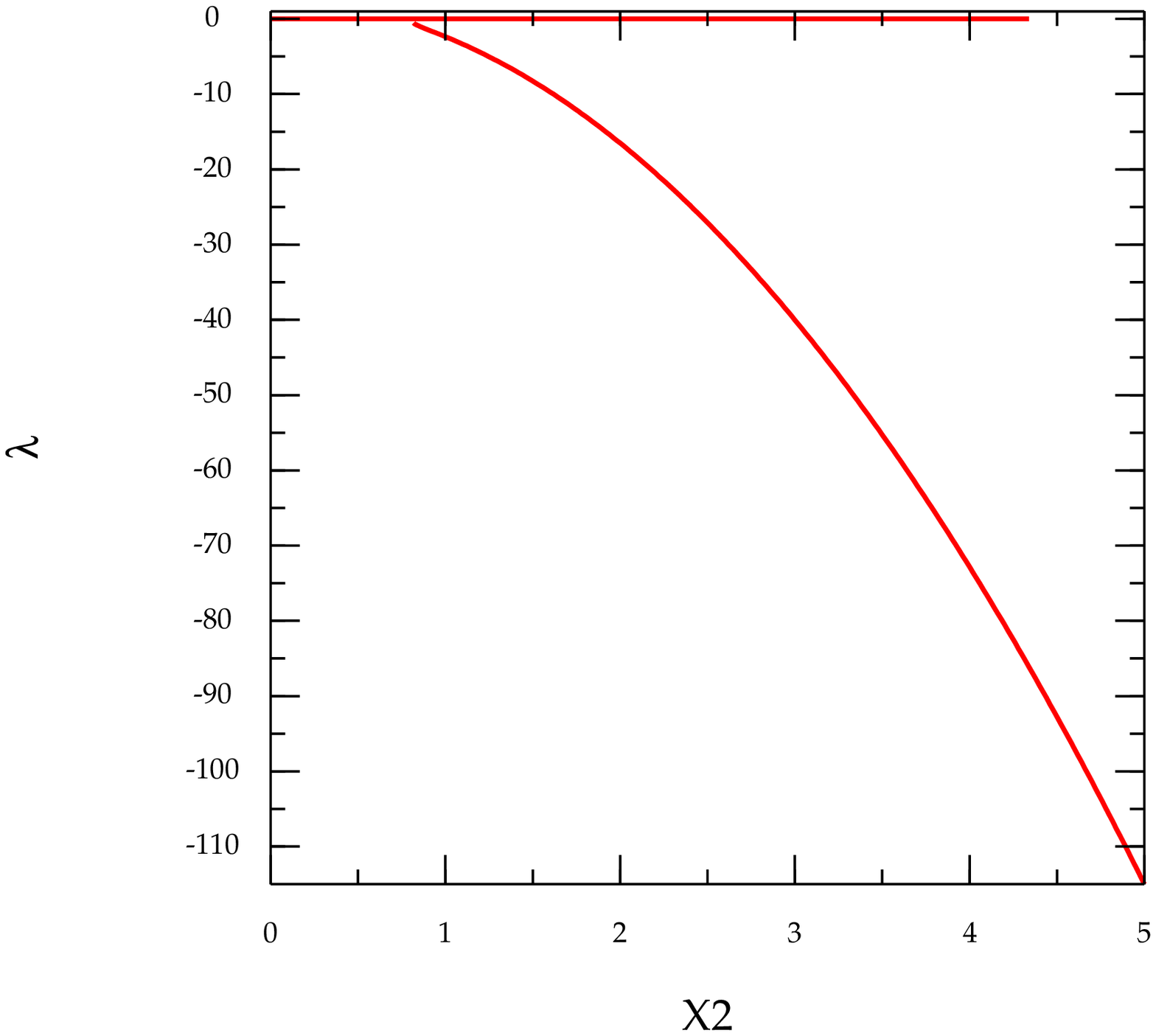}
\end{picture}
\caption{\label{X2n20}
The maximum of probability, $D = \hbox{Max}_{(i,j,\rho)}\,|\varphi_{i,j,\rho}|^2$ 
and the eigenvalue $\Lambda$ for $X_1=0.6$, $G=0.6$ 
as a function of $X_2$ for $n=20$.}
\end{figure}

In Fig. \ref{X2n20} we present the plots of the values of the maximum 
density and of the eigenvalues of
the solutions for  $X_1=0.6$, $G=0.6$ and $n=20$ as a function of $X_2$.
We note that in this case all solutions fall into two classes
seen before. The solutions with a small 
value of the maximum density are of type I while the highly peaked ones are of 
type VI.
The solutions look very much the same for other values of $n$ except that the
ranges of parameters for which the solutions exist do depend on $n$. 

\begin{figure}[htbp]
\unitlength1cm \hfil
\begin{picture}(8,8)
 \epsfxsize=8cm \epsffile{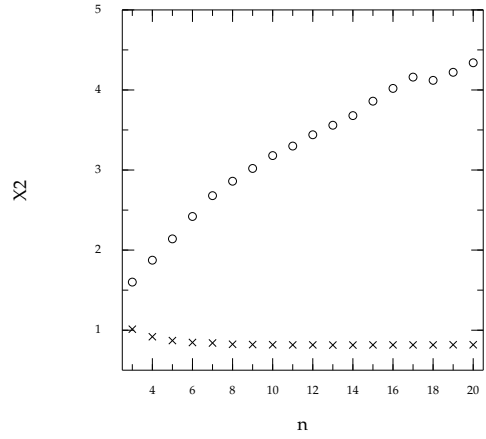}
\end{picture}
\caption{\label{TransX2}
Domain of solutions as a function of $n$ for solutions with
$X_1=0.6$, $G=0.6$. 
$\circ$: upper value of $X_2$ for ring-like solutions (type I). 
$\times$: lower value of $X_2$ for solutions of type VI.}
\end{figure}

In Fig. \ref{TransX2} we present the regions of parameters for which the 
solutions of type I 
and type VI exist as a function of $n$ and $X_2$. Solutions of type I exist
for all values of $X_2$ below the circles while those of type VI exist for all
values of $X_2$ above the crosses. 

We see that there is a large overlap of the regions in which the solutions of 
the type I and VI coexist and that the region of this overlap grows with $n$.


\subsection{Solutions for $X_2=0$, $G=0.6$ and various values of $X_1$}
 
\begin{figure}[htbp]
\unitlength1cm \hfil
\begin{picture}(16,8)
 \epsfxsize=8cm \epsffile{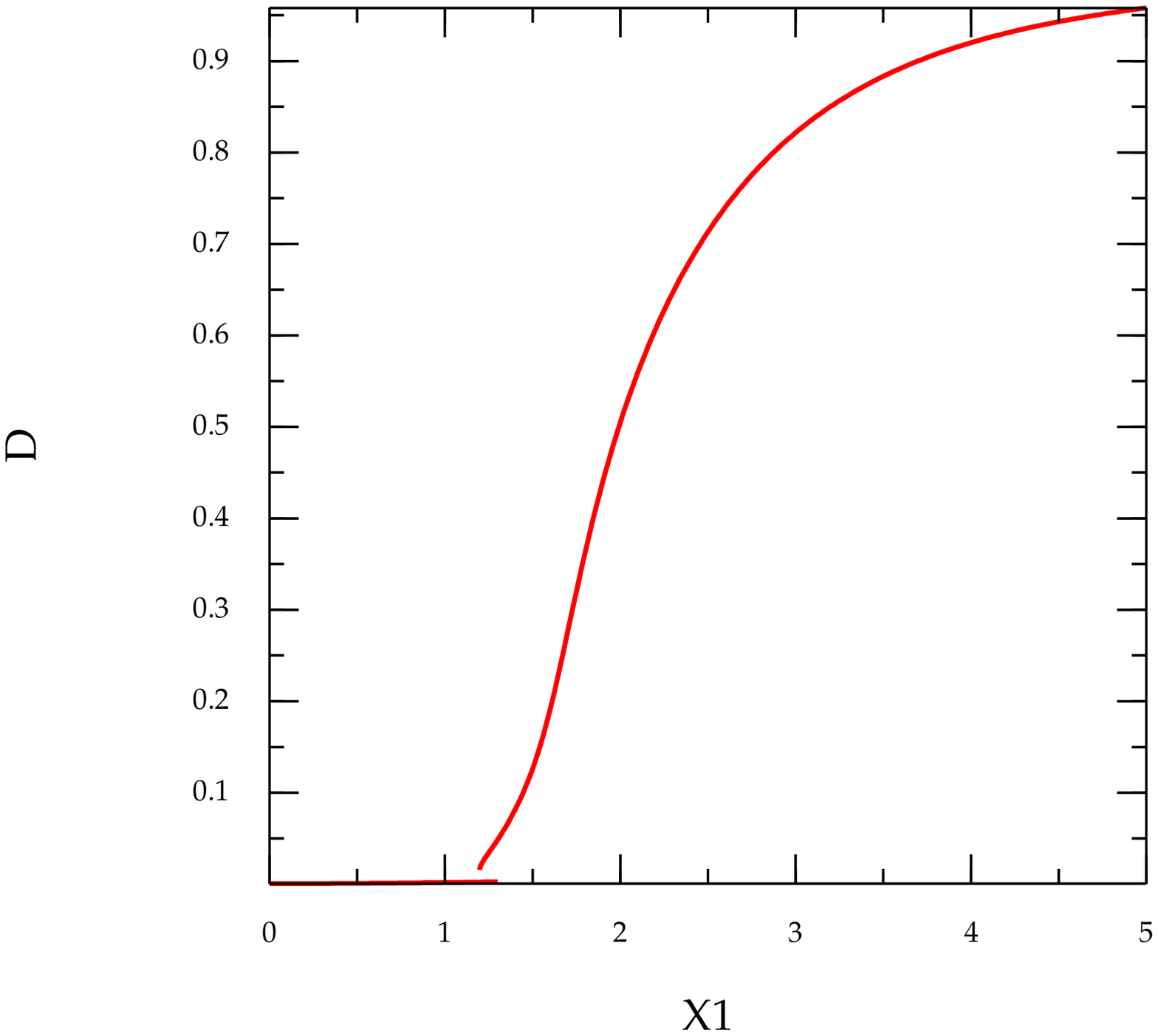}
 \epsfxsize=8cm \epsffile{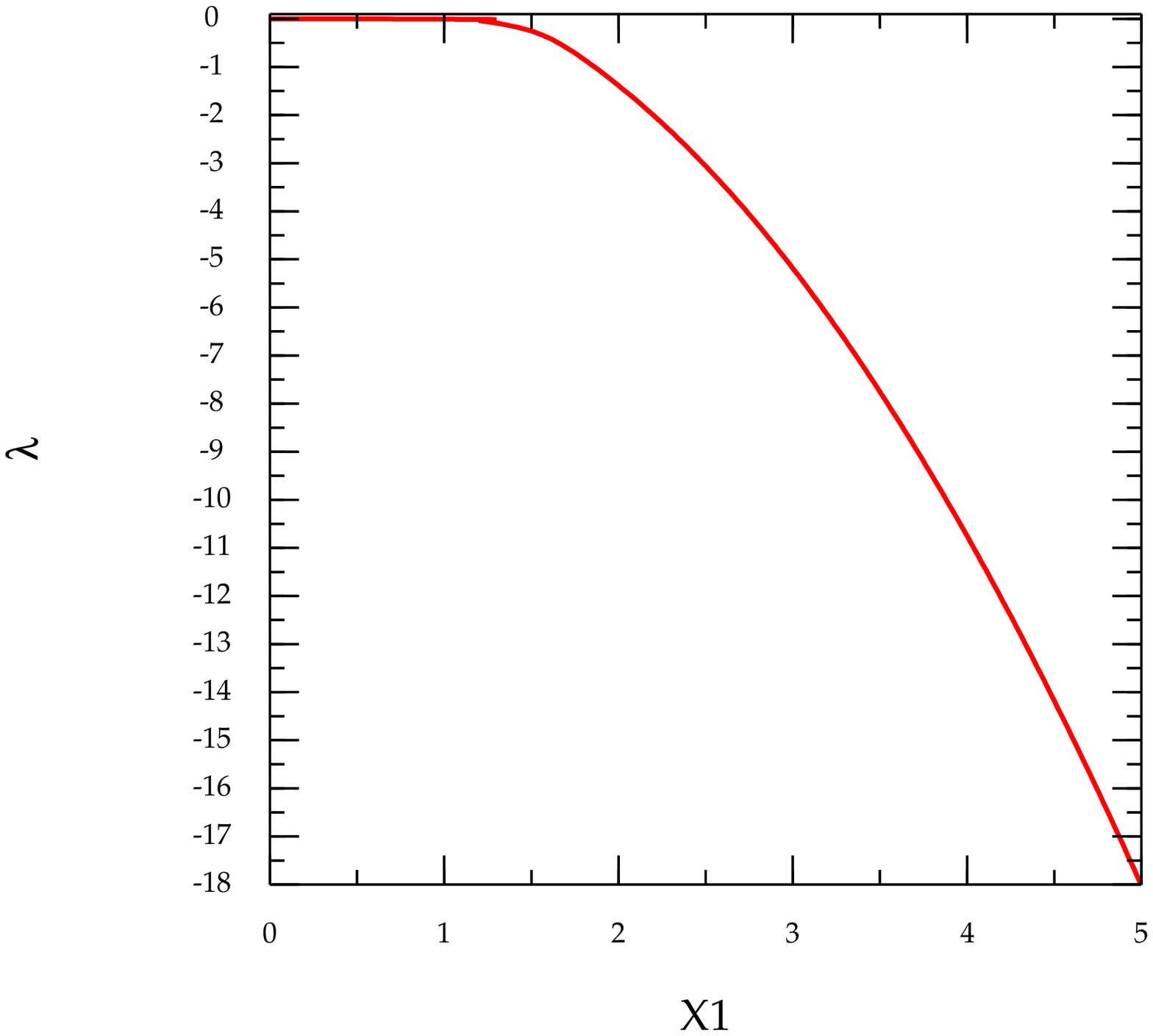}
\end{picture}
\caption{\label{X1n20}
The maximum of probability, $D = \hbox{Max}_{(i,j,\rho)}\,|\varphi_{i,j,\rho}|^2$ 
and the eigenvalue for $X_2=0$, $G=0.6$ as a function of $X_1$ for $n=20$.}
\end{figure}

In Fig. \ref{X1n20} we present the plots of the maximum density and of the 
eigenvalue of the solutions for  $X_2=0$, $G=0.6$ and $n=20$ as a function of 
$X_1$.
Again we note that there are two types of solutions in this case. The 
solutions with a small density maximum are of type I while 
the highly peaked ones are of type V. This time, however, there is very little 
overlap between these two classes of solutions.

The solutions look very much the same for other values of $n$, except that, 
again, the range of parameters where the concrete solutions exist does depend 
on $n$. For a given value of the 
parameters,  the solutions are also more bound when $n$ is small.

From this numerical study and what we know about the physical values of the
parameters in our model we can conclude that the type I solutions  
exist in all real nanotubes. If $X_1$ and $X_2$ are large, then strongly
localised solutions of other types may exist as well.

\section{Conclusions}

In this paper we have  determined the conditions for  the existence of polaron 
(narrow soliton) states in hexagonal nanotube systems. They are given 
by the solutions of the equations minimising the total Hamiltonian of the 
system.
We have also studied the properties of these solutions showing 
that in some cases the system possesses two or more different states  
very close in energy. Some of these states are localised on few lattice 
sites, 
some are more spread out. We have determined the boundaries
between various states and have found that in some cases these boundaries
are very sharp. The corresponding critical values of the coupling constants 
depend on the diameter of a nanotube. We have also shown that the solutions 
corresponding to the physical parameter values of the
carbon nanotube are ring-like solutions wrapped 
around the nanotube with a profile resembling that of the nonlinear 
Schr\"odinger 
soliton, {\it i.e.} similar to a 1D soliton (comp. \cite{Alves}).

Our numerical results do not apply straightforwardly to real carbon nanotubes 
with one electron per carbon atom when half of the electronic band states are 
occupied. Nevertheless, our results may be applicable to doped 
nanotubes at doping levels such that the concentration of charge carriers  
is sufficiently low so that the distances between self-trapped carriers exceed 
their localisation width. As has been shown in \cite{BrEr-Pauli}, in the 
case of the self-trapping, the Fermi statistics manifests itself by 
the appearance of the spatial repulsion of bipolarons. Two electrons with 
opposite spins can occupy the same place forming a bipolaron state. Two 
electrons with parallel spins or four electrons, in general, can create two 
self-trapped states.
These states are characterised by a Fermi repulsion  between them
which decreases exponentially with the increases of the distance between 
them \cite{BrEr-Pauli}. When the average distance between (bi)polarons
is larger than the localisation region, the interaction between them can be
neglected, and the single-(bi)polaron model, used in the present paper, 
becomes justified.

\section{Acknowledgement}
This work has been supported by a Royal Society travel grant.

{}

\end{document}